\documentclass[10pt,aps,prl, nofootinbib,superscriptaddress,reprint,tightenlines]{revtex4-1}
\usepackage{amssymb,amsmath,accents,mathrsfs}
\usepackage{verbatim,graphics,graphicx,color,slashed,textcomp,bbm,mathdots,multirow,array}
\usepackage{aas_macros}
\usepackage{graphicx}
\usepackage{color,units}
\usepackage{xcolor}
\usepackage{hyperref}
\usepackage{bm}
\usepackage[title]{appendix}  
\usepackage{listings}
\usepackage{soul} 
\usepackage{tabularx}
\usepackage[normalem]{ulem}
\hypersetup{
bookmarks=true,
unicode=false,
pdftoolbar=true,
pdfmenubar=true,
pdffitwindow=false,
pdfstartview={FitH},
pdfauthor={}
colorlinks=true,
linkcolor=blue,
citecolor=red,
urlcolor=blue,
}
\graphicspath{{}}

\begin{document}

\definecolor{dkgreen}{rgb}{0,0.6,0}
\definecolor{gray}{rgb}{0.5,0.5,0.5}
\definecolor{mauve}{rgb}{0.58,0,0.82}

\lstset{frame=tb,
  	language=Matlab,
  	aboveskip=3mm,
  	belowskip=3mm,
  	showstringspaces=false,
  	columns=flexible,
  	basicstyle={\small\ttfamily},
  	numbers=none,
  	numberstyle=\tiny\color{gray},
 	keywordstyle=\color{blue},
	commentstyle=\color{dkgreen},
  	stringstyle=\color{mauve},
  	breaklines=true,
  	breakatwhitespace=true
  	tabsize=3
}

\title{Detecting ultralight dark matter in the Galactic Center with pulsars around Sgr~A*}

\author{Jiang-Chuan Yu}
\affiliation{Department of Astronomy, School of Physics, Peking University, Beijing 100871, China}
\affiliation{Kavli Institute for Astronomy and Astrophysics, Peking University, Beijing 100871, China}
\author{Yan Cao}
\affiliation{School of Physics, Nanjing University, Nanjing 210093, China}
\author{Zexin Hu}
\affiliation{Department of Astronomy, School of Physics, Peking University, Beijing 100871, China}
\affiliation{Kavli Institute for Astronomy and Astrophysics, Peking University, Beijing 100871, China}
\author{Lijing Shao}
\email{lshao@pku.edu.cn}
\affiliation{Kavli Institute for Astronomy and Astrophysics, Peking University, Beijing 100871, China}
\affiliation{National Astronomical Observatories, Chinese Academy of Sciences, Beijing 100012, China}
\date{\today}

\begin{abstract}
Ultralight dark matter (ULDM) model is a leading dark matter candidate that arises naturally in extensions of the Standard Model. In the Galactic Center, ULDM manifests as dense hydrogen-like boson clouds or self-gravitating soliton cores. We present the first study of the gravitational effects of these ULDM structures on pulsar orbits around Sgr~A*, using pulsar timing as a precision dynamical probe, based on a comprehensive and practical framework that includes various kinds of black hole and orbital parameters.
Our analysis shows that long-term pulsar monitoring—one of the key objectives of future SKA science—could detect a boson cloud with a total mass as low as $\mathcal{O}(M_\odot)$ for boson mass $m \sim 10^{-18}\,\mathrm{eV}$, and probe a wide range of soliton core masses in the lower-mass regime, assuming a conservative timing precision of $\sigma_{\mathrm{TOA}}=1\,\mathrm{ms}$.

\end{abstract}

\maketitle

\textit{Introduction.}---Ultralight bosons with mass $m \lesssim 10^{-6}\,\mathrm{eV}$ have emerged as compelling dark matter candidates, since they naturally alleviate challenges on the galactic scale in the traditional cold dark matter paradigm~\cite{Hu:2000ke,DelPopolo:2016emo,Hui:2016ltb,Ferreira:2020fam,Singh:2025uvp} and are predicted by various theories beyond the Standard Model~\cite{Peccei:1977hh,Weinberg:1977ma,Wilczek:1977pj,Hui:2016ltb,Marsh:2015xka,Agrawal:2021dbo,Polchinski:1998rr,Kim:1986ax,Fayet:1990wx,Svrcek:2006yi,Arvanitaki_2010,Arias:2012az,Lyth:1998xn,PhysRevD.93.103520,Ema:2019yrd,Ahmed:2020fhc,Kolb:2020fwh}. Independent constraints on ultralight dark matter (ULDM) arise from both cosmological and local measurements. Stellar kinematics in dwarf galaxies~\cite{Bar:2018acw,Marsh:2018zyw,Pozo:2023zmx,Pozo:2020ukk,PhysRevLett.134.151001,may2025updatedboundsultralightdark} and other substructures~\cite{Church:2018sro,Amorisco:2018dcn} favors solitonic cores formed by ULDM with boson mass $m \gtrsim 10^{-22}\,\mathrm{eV}$. Observations of the Lyman-$\alpha$ forest~\cite{Armengaud:2017nkf,PhysRevLett.119.031302,Zhang:2017chj,Kobayashi:2017jcf} and 21\,cm absorption line~\cite{Schneider:2018xba,Lidz:2018fqo} further limit the particle mass to $m \gtrsim 10^{-21}\,\mathrm{eV}$. Precision experiments, such as pulsar timing observations~\cite{Khmelnitsky:2013lxt,Porayko:2014rfa,Nomura:2019cvc,Sun_2022,Dror:2025nvg} and laser interferometers~\cite{Aoki:2016kwl,Kim:2023pkx,PhysRevD.110.023025} can also constrain oscillating ULDM fields via their coupling to the metric. Specifically, pulsar timing array experiments~\cite{EuropeanPulsarTimingArray:2023egv} showed that only ULDM with $m \gtrsim 10^{-23}\,\mathrm{eV}$ can constitute all of the local dark matter density.



Our Galactic Center (GC) provides a natural laboratory to probe the properties of dark matter~\cite{Gondolo_1999,Sadeghian:2013laa,Hu:2023ubk,Cheng:2024mgl,GRAVITY:2019tuf,Yuan:2022nmu,GRAVITY:2023cjt,GRAVITY:2023azi,Bar:2019pnz,Lacroix:2018zmg,Chan_2022,Shen:2023kkm,Zakharov:2007fj,Heissel:2021pcw}, as dense dark matter structure can form around the supermassive black hole (SMBH). Stellar orbital motion, especially that of S2, has been widely used to constrain the dark matter distribution near Sgr~A*~\cite{Lacroix:2018zmg,Shen:2023kkm,Zakharov:2007fj,Heissel:2021pcw}. Bound structures of ULDM, such as superradiant cloud~\cite{PhysRevD.22.2323,Cardoso:2005vk,PhysRevD.76.084001,Arvanitaki:2010sy,PhysRevD.87.043513,Brito:2014wla,Brito:2015oca,Baryakhtar:2017ngi,Frolov:2018ezx,Dolan:2018dqv,Brito_2020,B2,B3,Siemonsen:2022yyf} and soliton core~\cite{Schive:2014dra,Schive:2014hza,Chavanis:2019bnu,Bar:2019pnz,Davies_2020,Annulli_2020,Zagorac_2023,Aghaie_2024,liao2025decipheringsolitonhalorelationfuzzy}, may exist in the GC. Recent studies~\cite{GRAVITY:2019tuf,Yuan:2022nmu,Chen_2023,GRAVITY:2023cjt,GRAVITY:2023azi,Chen_2025,bai2025probingaxionsspectroscopicmeasurements,tomaselli2025probingdenseenvironmentssgr,Bar:2019pnz,Chan_2022} have investigated their observational signatures and constrained the total mass of such structures. 

Pulsars are rapidly rotating neutron stars whose remarkably stable spin periods make them exquisite clocks for probing gravitational dynamics. Through precisely measuring times of arrival (TOAs) of their pulses, a pulsar in a tight orbit around Sgr~A*, potentially discoverable with next-generation instruments such as the Square Kilometre Array (SKA), would offer an unique opportunity to test gravity in the strong-field regime~\cite{Wex:1998wt,Liu:2011ae,Psaltis:2015uza,Zhang:2017qbb,Bower:2018mta,Dong:2022zvh,Hu:2023ubk,Hu:2023vsg,Hu:2024blq,Shao:2025vmb} and explore the dark matter properties in the Galactic Center~\cite{DeMartino:2017qsa,Hu:2023ubk}.


In this {\it Letter}, we assess the detectability of nonrelativistic ULDM structures in the GC, including the gravitational atom (GA) and self-gravitating spherical soliton, through the timing observation of a pulsar with a realistic orbital period $P_b\in [0.5,5]\,\mathrm{yr}$. We develop a post-Newtonian timing framework for the pulsar–SMBH binary incorporating the leading-order gravitational perturbation induced by the ULDM struture and estimate the sensitivity of long-term observations to its total mass. Throughout this {\it Letter}, we adopt the natural units where $\hbar=G=c=1$.

\textit{ULDM in the GC.}---If the boson is sufficiently light, the ULDM environment of Sgr~A* locating far away from the BH horizon can be modeled as a nonrelativistic structure. In this limit, the slow mode of a massive spin-$s$ bosonic field is described by a rank-$s$ tensor wavefunction $\psi_I$ in the Cartesian basis, where $I$ refers schematically to the set of spatial tensor indices (e.g., $\psi$ for scalar, $\psi_i$ for vector) \cite{Baryakhtar:2017ngi,Brito_2020,Jain:2021pnk,Cao:2023fyv,Cao:2024wby}  . Neglecting the possible non-gravitational self-interaction, the dynamics of $\psi_I$ around a static point mass $M$ is governed by the Schr\"odinger-Poisson (SP) equation,
\begin{align}
i\partial_t\psi_I &= -\frac{1}{2m}\nabla^2\psi_I + m\left(\Phi - \frac{M}{r} \right)\psi_I, \\
\nabla^2\Phi &= 4\pi \rho,\quad \rho = m \sum_I |\psi_I|^2,
\end{align}
where $\rho$ and $\Phi$ are the mass density and Newtonian potential of the wavefunction respectively. We focus on two classes of bound states: the GA for which $\Phi$ is negligible, and the spherically symmetric ground state referred to as \textit{spherical soliton}.

The total mass of the bound state is given by $
M_\mathrm{c} = \int d^3r\, m \sum_I |\psi_I|^2 \equiv \beta M$. If the mass ratio $\beta \ll 1$, self-gravity of the wavefunction can be neglected at the leading order. Setting $\Phi=0$, the SP equation reduces to the Shr\" {o}dinger equation of a hydrogen atom, with the gravitational fine-structure constant $\alpha\equiv m M\approx 0.032(\mu/10^{-18}\text{eV})(M/4.3\times 10^6M_\odot)$ and Bohr radius $r_\text{c}\equiv M/\alpha^2$. Consequently, $\psi_I$ is spanned by the hydrogenic bound states $\psi^{(nl\texttt{m})}\propto R_{nl}(r)\,Y_{l\texttt{m}}(\theta,\phi)$ for scalar field, labeled by the quantum numbers $|nl\texttt{m}\rangle$; in the case of vector field, the angular wavefunction is spanned by the pure-orbital vector spherical harmonics, $\mathbf{Y}_{lj\texttt{m}}(\theta,\phi)$, and an eigenstate can thus be labeled by $|nlj\texttt{m}\rangle$ (see \cite{SupplementaryMaterials} for details). For $\alpha \ll 1$, since $r_\text{c}\gg M$, this GA model provides an effective description for the superradiant cloud (quasibound states) around a spinning BH at radius $r\gg M$, with the $z$-axis aligned with the BH’s spin direction. The rotational superradiant instability varies between different states \cite{B2,B3}. In this work, we consider the GA occupied by the fastest growing state for an initial BH spin $\chi_i\sim 1$ (the so-called superradiant ground state), which is the $|211\rangle$ state for scalar field, the $|1011\rangle$ state for vector field, and the $|1022\rangle$ state for spin-2 field. The density profiles of the latter two are degenerate with that of the scalar $|100\rangle$ state. Neglecting the baryonic accretion, the saturated cloud mass is given by $\beta_\text{max}\approx \alpha \chi_i/\texttt{m}$, thus $\beta\ll 1$ is automatically satisfied.

The GA approximation breaks down when $\beta$ becomes sufficiently large, which is typical for a soliton core; in this regime, the bound state should be computed nonperturbatively. We consider the spherically symmetric ground state of the SP equation in the form $\psi(t,r)=f(r)\,e^{-iE t}$, which describes a nodeless self-gravitating configuration around the central mass. Due to a scaling symmetry, the SP equation for given $\alpha$ admits solutions of a one-parameter family~\cite{SupplementaryMaterials} in the dimensionless coordinate $y\equiv m r$, labeled by $\kappa^2\equiv \sqrt{{8\pi}/{m}}\,f(0)$. For $\beta=0$, these solutions reduce to the $|100\rangle$ state of the scalar GA, and for a moderately small $\beta$, the ground state is still well-described by a $|100\rangle$ state corrected by its own Newtonian potential, with the energy level $E = -(m\alpha^2/2)\,C(\beta)$ and $C(\beta)\approx 1+5\beta/4$. As the soliton mass increases, the configurations become more compact, and relativistic corrections are correspondingly more important \cite{Salehian_2021,zhang2025unifiedviewscalarvector}. We thus assume that $r_{0.5}\ge M+M_\text{c}$, with $r_{0.5}$ being the radius enclosing 50\% of the total soliton mass. This condition turns out to be well satisfied in the parameter space of interests.

\begin{figure}
\includegraphics[scale=0.25]{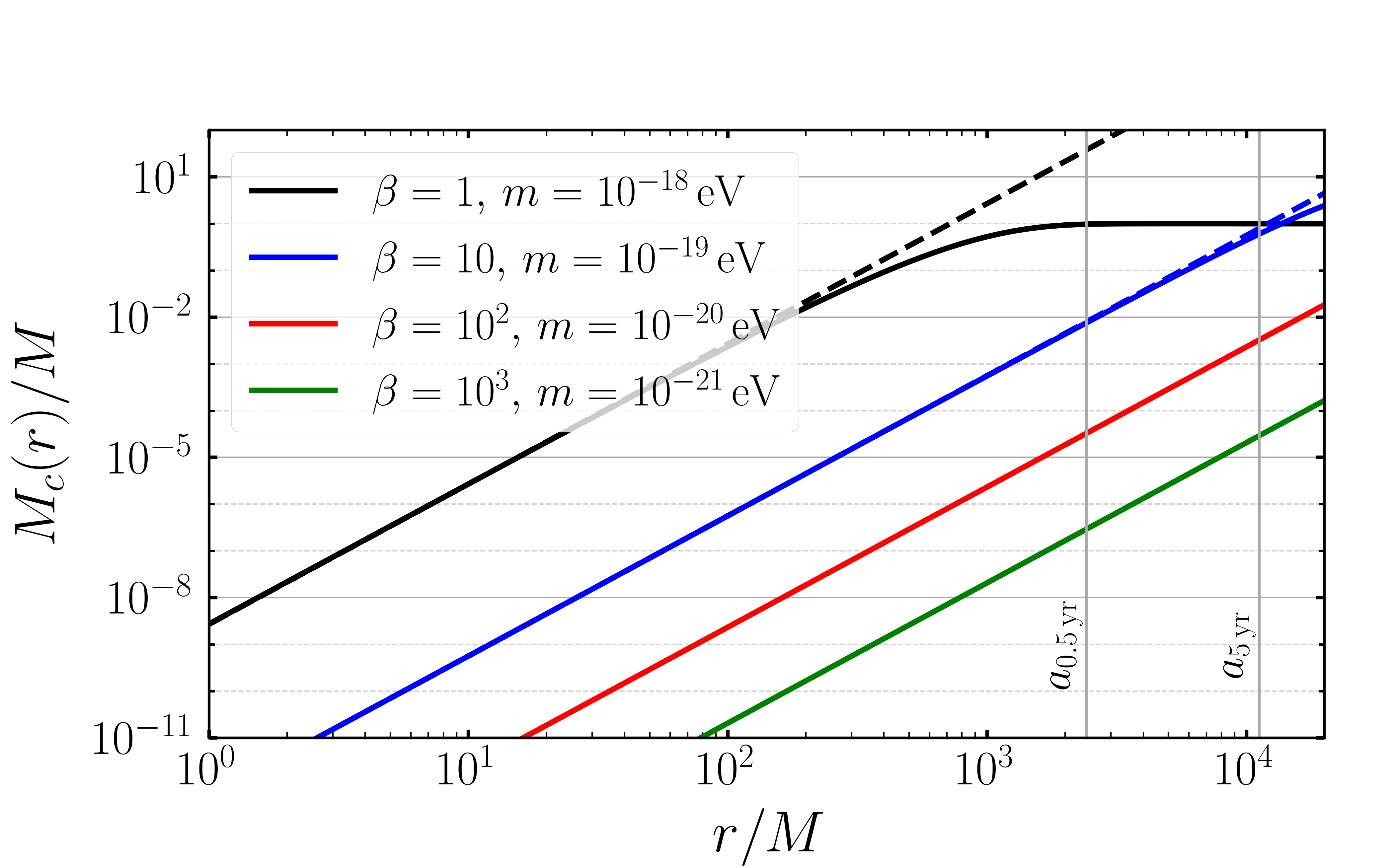}
\caption{Enclosed mass of the spherical soliton in the GC for various boson masses and soliton masses.  The inner profiles are well approximated by $m^6\,D(\beta)\,(Mr)^3/6$, shown as dashed lines. The semi-major axes of the considered pulsar orbits with given periods are indicated by the gray vertical lines. The mass of Sgr~A* is $M=4.3\times 10^6M_\odot$.}
\label{soliton_mass_profile}
\end{figure}

The normalized density profile depends only on $\beta$ and $r/r_\text{c}$, hence for given $\beta$, a larger boson mass results in a more compact soliton. This is reflected in the flat density profile of the inner region~\cite{SupplementaryMaterials}: $M_\text{c}(r)/M\approx (4\pi/3)\,\rho(0)\, r^3/M=m^6\,D(\beta)\,(Mr)^3/6$, with $D(\beta)\approx 0.11 \beta^4$ for $\beta \gtrsim 100$, as depicted in Fig.~\ref{soliton_mass_profile}. For a large soliton mass, the enclosed mass within $r=100M$ in the GC is therefore $M_\text{c}(r)/M_\text{c}\approx 1.8\times 10^4\,\beta^3\alpha^6\approx 0.02\,(\mu/10^{-19}\text{eV})^6(\beta/1000)^3$, thus most of the soliton mass would indeed locate far away from the BH if $\beta \lesssim 1000\,(\mu/10^{-19}\text{eV})^{-2}$, in which case the accretion of the soliton onto the BH is suppressed~\cite{Annulli_2020}.

\textit{Forecast of sensitivities.}---To quantify the sensitivity of a single pulsar to the aforementioned ULDM structures in the GC, we simulate the timing signals using the numerical framework developed in Ref.~\cite{Hu:2023ubk}. The orbital evolution is obtained by integrating the post-Newtonian equations of motion, with the DM-induced acceleration included as $\ddot{\mathbf{r}}_{\text{DM}}\approx -\nabla \Phi$, since the dissipative effects and relativistic corrections from ULDM are negligible \cite{Cao:2024wby}.


The model parameters are grouped into three sectors: the black hole parameters $\Theta_{\text{BH}} = \{M, \chi, q, \lambda, \eta\}$, describing the mass, spin, quadrupole moment, and spin orientation; the pulsar orbital and spin parameters $\Theta_{\text{PSR}} = \{P_b, e, \omega, i, \theta_0, N_0, \nu, \dot{\nu}\}$, where $\theta_0$ is the initial orbital phase and ${N_0,\nu,\dot \nu}$ is associated to the rotation number $N(T)=N_0+\nu T+\dot \nu T^2/2$ as function of proper time $T$; and the ULDM parameters $\Theta_{\text{DM}} = \{\beta\}$. For simplicity, we fix the unobservable longitude of the ascending node to $\Omega=0$. Timing residuals are computed including the Römer, Shapiro, and Einstein delays, and the effects due to the SMBH’s proper motion are also included. Additional numerical details are provided in \cite{SupplementaryMaterials}.

We employ the Fisher matrix formalism to estimate the projected constraints on ULDM parameters. The covariance matrix is given by $C_{\mu\nu} = \left(\partial^2 \mathcal{L}/\partial\Theta^\mu \partial\Theta^\nu\right)^{-1}$, where the log-likelihood takes the form
\begin{equation}
\mathcal{L} = \frac{1}{2\nu^2} \sum_{i=1}^{N_{\mathrm{TOA}}} \frac{\left[N_i(\Theta) - N_i(\bar{\Theta})\right]^2}{\sigma_{\mathrm{TOA}}^2},
\end{equation}
with $N_i(\Theta)$ denoting the pulsar rotation number at the $i$-th TOA $t^{\mathrm{TOA}}_i$, and $\bar{\Theta}$ the true system parameters. We adopt a timing precision of $\sigma_{\mathrm{TOA}} = 1\,\mathrm{ms}$ and assume a total observation time of $T_{\mathrm{obs}} = 5\,\mathrm{yr}$, consistent with the projected capabilities of the Square Kilometre Array (SKA) and next-generation Very Large Array (ngVLA).

\begin{figure}[htbp]
    \centering
    \includegraphics[scale=0.25]{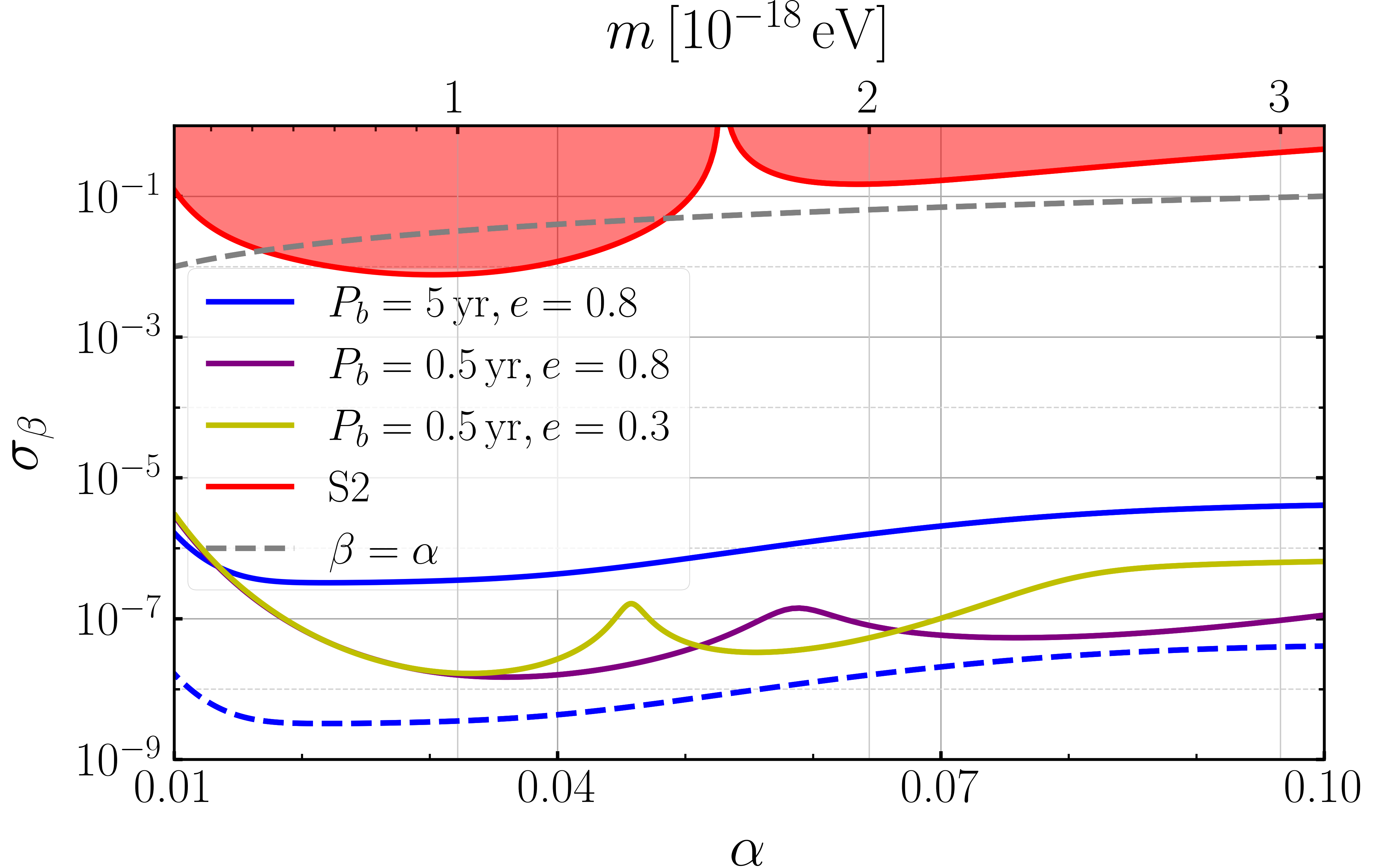}\\[0.3cm]  
    \includegraphics[scale=0.25]{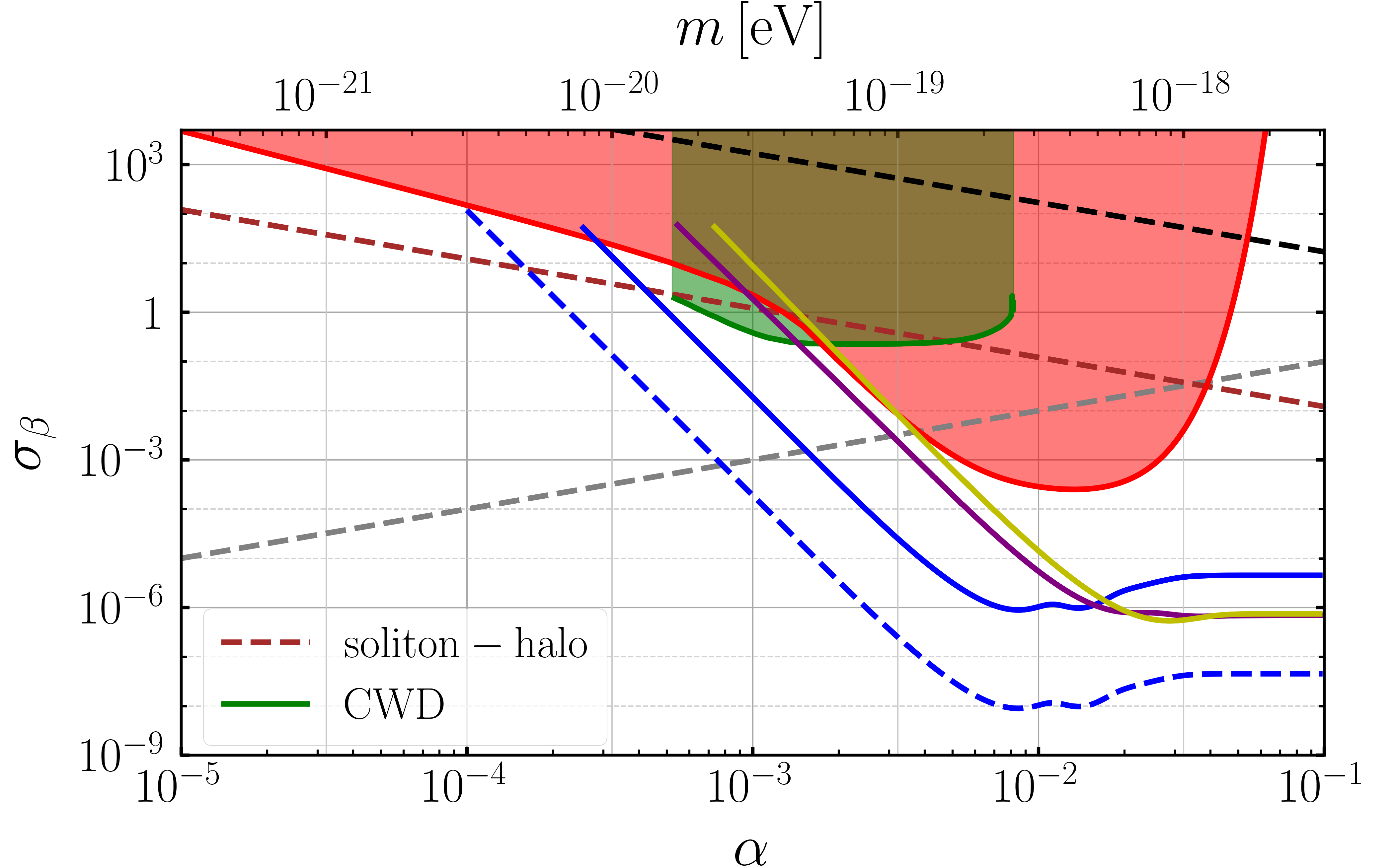}
    \caption{Projected sensitivities to the mass ratio $\beta$ versus $\alpha$ for scalar GA in the $|211\rangle$ state (top) and spherical soliton (bottom). The result for the soliton also applies to the vector or spin-2 GA in its superradiant ground state for $\beta<\alpha$. 
    Red region is excluded by Schwarzschild precession of S2 star and 
    the green region is excluded by stellar dynamics of the clockwise rotating disk~\cite{Beloborodov:2006is}. 
    The blue dashed line denotes the sensitivities in case of timing precision $\sigma_{\text{TOA}}=10\, \mu\text{s}$.
    The gray dashed line marks the boundary $\beta = \alpha$, corresponding to the threshold of superradiance. 
    Below the black dashed line, the soliton satisfies $r_{0.5}>M+M_\text{c}$. 
    The brown dashed line shows the soliton mass given by the soliton-halo relation for our galaxy~\cite{Schive:2014dra,Schive:2014hza}.
    }
    \label{fig:combined sensitivities}
\end{figure}


Figure~\ref{fig:combined sensitivities} presents the projected sensitivities to the mass ratio of the cloud or soliton core to the central black hole, $\beta$, for two benchmark profiles: the scalar $|211\rangle$ GA and the spherical soliton. Estimated constraints~\cite{SupplementaryMaterials} from the periastron precession of the S2 star~\cite{GRAVITY:2020gka} and the stellar dynamics of the clockwise rotating disk (CWD)~\cite{Beloborodov:2006is} are shown for comparison. Note that the $3\sigma$ constraints obtained by the GRAVITY collaboration \cite{GRAVITY:2023cjt,GRAVITY:2023azi}, $\beta_\text{211}\lesssim 10^{-3}$ for $\alpha\in(0.015,0.045)$ and $\beta_\text{1011} \lesssim 10^{-3}$ for $m\in(10^{-19},10^{-18})\,\text{eV}$, are slightly stronger than our $1\sigma$ estimations in the corresponding mass range. The brown dashed line shows the soliton mass given by the halo–soliton relation, which is extrapolated from the dark-matter-only simulations~\cite{Schive:2014dra,Schive:2014hza} in the case of scalar ULDM.



For the scalar $|211\rangle$ configuration, a pulsar on a 0.5-year orbit yields new sensitivity on the mass ratio $\beta \sim 10^{-7}$ in the range $10^{-2} \lesssim \alpha \lesssim 10^{-1}$, improving upon S2 constraints by up to five orders of magnitude. Notably, the orientation of the orbit relative to the BH's spin also introduces angular dependence, leading to an order-of-magnitude variation in the 
sensitivities; more details can be found in~\cite{SupplementaryMaterials}. For the spherical soliton, we find that a 0.5-yr pulsar orbit will place the strongest constraints on $\beta$ at $\alpha \gtrsim 10^{-3}$. The projected sensitivity extends to $\beta \sim 10^{-6}$ at high masses, improving significantly over current bounds, typically much helpful to search for superradiantly generated GAs ($\beta \lesssim \alpha$).
We can see that, in both ULDM scenarios, pulsars with long orbital periods tend to exhibit better sensitivities at smaller $\alpha$, whereas those with relatively short orbital periods tend to have better sensitivities at larger $\alpha$. For pulsars with the same period, those with a higher eccentricity perform better in the low-mass range. Importantly, our method can probe ULDM down to $\alpha \gtrsim 3\times10^{-4}$, surpassing the soliton mass scale inferred from the empirical halo–soliton relation. Moreover, this relation—extrapolated from galactic simulations—has been shown to overestimate the soliton mass by orders of magnitude for $m \gtrsim 10^{-21}$\,eV~\cite{Bar:2019pnz}, further highlighting the advantage of pulsar-based timing over traditional stellar dynamics in the GC. Moreover, if the timing precision reaches $\sigma_{\text{TOA}}=10 \mu$s in the future, the corresponding sensitivities would be 100 times better, as shown in Fig.~\ref{fig:combined sensitivities}.

The size of GA (with $\beta\ll 1$) is characterized by the Bohr radius $r_\text{c}=M/\alpha^2$. As a result, when the pulsar is within the Bohr radius, the enclosed mass and hence the gravitational influence of GA roughly increases with the size of the orbit. For a given pulsar orbit, this leads to enhanced sensitivity with increasing $\alpha$. The situation changes for orbits far outside the Bohr radius, due to an exponential decay in mass density with radius. The size of the spherical soliton for a given $\beta$ is also controlled by the Bohr radius. Although parametrically smaller than that of GA with the same $\alpha$, in the small-$\alpha$ regime the soliton size remains much larger than the considered pulsar orbits for the marginally sensitive values of $\beta$ in Fig.~\ref{fig:combined sensitivities}, leading to a decrease in sensitivity with decreasing $\alpha$.

In practice, the result of parameter estimation is also affected by other factors. For example, the presence of the black hole's spin and quadrupole moment can partially degenerate with the dark matter effect, degrading the sensitivities on $\beta$ by $\sim \mathcal{O}(1)-\mathcal{O}(10)$. However, our results show no significant dependence on the assumed spin magnitude of the black hole.




\textit{Summary and discussions.}---We investigate the potential of using a pulsar near the GC to probe ULDM through pulsar timing observations. Assuming a pulsar with an orbital period \( P_b \sim 0.5\,\mathrm{yr} \), eccentricity \( e \sim 0.8 \), and timing precision $\sigma_{\text{TOA}}=1$ ms over a 5-yr observation with weekly TOA measurements, simulations show that with ULDM masses in the range \( 3\times10^{-19}\,\mathrm{eV} < m < 3\times10^{-18}\,\mathrm{eV} \), the mass ratio of the cloud or soliton core to the central black hole $\beta$ can be constrained to amplitudes of order \( \mathcal{O}(10^{-7}\text{--}10^{-6}) \), improving upon existing limits from S2 star at least by a few orders of magnitude. In the case of a spherical soliton, a  pulsar with $P_b = 5\,{\rm yr}$ provides excellent sensitivity in the range of $5 \times 10^{-4} \lesssim \alpha \lesssim 10^{-2}$, significantly exceeding the total mass predicted by the empirical halo--soliton relation. These measurements improve proportionally if better timing precision is achieved.


In the mass window $0.066\lesssim \alpha \lesssim 0.16$ (scalar) and $0.015\lesssim \alpha \lesssim 0.039$ (vector), the timescale of the superradiant growth of GA for an initial BH spin $\chi_i\sim 1$ is shorter than the characteristic merger time ($\sim 10^{6}\,\mathrm{yr}$), while the saturated GA depletes via gravitational radiation on a timescale $\gtrsim 10^{8}\,\mathrm{yr}$. In the absence of other processes, the cloud would thus remain close to its saturated state (with $\beta \sim \alpha$), and future pulsar timing observations could either reveal its presence or place stringent bounds on the boson mass. In the more general scenario, the GA can be occupied by multiple states \cite{Ficarra_2019,Siemonsen_2020,Guo_2023,guo2025effectaccretionscalarsuperradiant} (see also \cite{SupplementaryMaterials}) and the soliton core will not be in a perfectly spherically symmetric ground state \cite{Zagorac_2022,Glennon_2023,salasnich2025collectiveexcitationsselfgravitatingultralight}. In such cases, the interference in the density profile can lead to oscillating gravitational potential and possibly distinctive orbital signatures. Additionally, relativistic effects would become important if the ULDM structure is sufficiently compact or resides closer to the BH horizon. These issues deserve further investigation.


We have not taken into account the corrections to the Shapiro and Einstein delays due to the gravitational potential of the ULDM structure, which are estimated to be negligible compared with the correction to the Römer delay. According to the model of \cite{Khmelnitsky:2013lxt,DeMartino:2017qsa}, the metric perturbation sourced by the fast mode of the bosonic field gives a timing residue of order $\rho/m^3$, this is much smaller than $\sigma_{\text{TOA}} = 1 \, \text{ms}$ in the interested parameter space \cite{SupplementaryMaterials} and consequently can also be neglected.

Finally, throughout this work, we have focused on the gravitational signature of ULDM in pulsar orbits near Sgr~A$^*$, while neglecting environmental perturbations such as stellar clusters~\cite{PhysRevD.81.062002,Liu_2012}, gas dynamics, and other external gravitational fields, as well as their impact on the ULDM structure~\cite{Du_2022}. These environmental effects are expected to be subdominant for tight orbits and, while they may introduce noise or systematics in pulsar timing, a full assessment of their impact on ULDM sensitivity requires more realistic astrophysical modeling.

\begin{acknowledgments}
This work was supported
by the National Natural Science Foundation of China (12573042), the National SKA Program of China (2020SKA0120300), the Beijing Natural
Science Foundation (1242018), 
the Max Planck Partner Group Program funded by the
Max Planck Society, and the High-Performance Computing Platform of Peking
University.
\end{acknowledgments}

\bibliography{paper}{}

\begin{thebibliography}{150}
\providecommand{\natexlab}[1]{#1}
\providecommand{\url}[1]{\texttt{#1}}
\expandafter\ifx\csname urlstyle\endcsname\relax
  \providecommand{\doi}[1]{doi: #1}\else
  \providecommand{\doi}{doi: \begingroup \urlstyle{rm}\Url}\fi

\bibitem[Hu et~al.(2000)Hu, Barkana, and Gruzinov]{Hu:2000ke}
Wayne Hu, Rennan Barkana, and Andrei Gruzinov.
\newblock {Cold and fuzzy dark matter}.
\newblock \emph{Phys. Rev. Lett.}, 85:\penalty0 1158--1161, 2000.
\newblock \doi{10.1103/PhysRevLett.85.1158}.

\bibitem[Del~Popolo and Le~Delliou(2017)]{DelPopolo:2016emo}
Antonino Del~Popolo and Morgan Le~Delliou.
\newblock {Small scale problems of the $\Lambda$CDM model: a short review}.
\newblock \emph{Galaxies}, 5\penalty0 (1):\penalty0 17, 2017.
\newblock \doi{10.3390/galaxies5010017}.

\bibitem[Hui et~al.(2017)Hui, Ostriker, Tremaine, and Witten]{Hui:2016ltb}
Lam Hui, Jeremiah~P. Ostriker, Scott Tremaine, and Edward Witten.
\newblock {Ultralight scalars as cosmological dark matter}.
\newblock \emph{Phys. Rev. D}, 95\penalty0 (4):\penalty0 043541, 2017.
\newblock \doi{10.1103/PhysRevD.95.043541}.

\bibitem[Ferreira(2021)]{Ferreira:2020fam}
Elisa G.~M. Ferreira.
\newblock {Ultra-light dark matter}.
\newblock \emph{Astron. Astrophys. Rev.}, 29\penalty0 (1):\penalty0 7, 2021.
\newblock \doi{10.1007/s00159-021-00135-6}.

\bibitem[Singh et~al.(2025)Singh, Brando~de Oliveira, Savastano, and
  Zumalac{\'a}rregui]{Singh:2025uvp}
Shashwat Singh, Guilherme Brando~de Oliveira, Stefano Savastano, and Miguel
  Zumalac{\'a}rregui.
\newblock {Gravitational wave lensing: probing Fuzzy Dark Matter with LISA}.
\newblock \emph{JCAP}, 07:\penalty0 025, 2025.
\newblock \doi{10.1088/1475-7516/2025/07/025}.

\bibitem[Peccei and Quinn(1977)]{Peccei:1977hh}
R.~D. Peccei and Helen~R. Quinn.
\newblock {CP Conservation in the Presence of Instantons}.
\newblock \emph{Phys. Rev. Lett.}, 38:\penalty0 1440--1443, 1977.
\newblock \doi{10.1103/PhysRevLett.38.1440}.

\bibitem[Weinberg(1978)]{Weinberg:1977ma}
Steven Weinberg.
\newblock {A New Light Boson?}
\newblock \emph{Phys. Rev. Lett.}, 40:\penalty0 223--226, 1978.
\newblock \doi{10.1103/PhysRevLett.40.223}.

\bibitem[Wilczek(1978)]{Wilczek:1977pj}
Frank Wilczek.
\newblock {Problem of Strong $P$ and $T$ Invariance in the Presence of
  Instantons}.
\newblock \emph{Phys. Rev. Lett.}, 40:\penalty0 279--282, 1978.
\newblock \doi{10.1103/PhysRevLett.40.279}.

\bibitem[Marsh(2016)]{Marsh:2015xka}
David J.~E. Marsh.
\newblock {Axion Cosmology}.
\newblock \emph{Phys. Rept.}, 643:\penalty0 1--79, 2016.
\newblock \doi{10.1016/j.physrep.2016.06.005}.

\bibitem[Agrawal et~al.(2021)]{Agrawal:2021dbo}
Prateek Agrawal et~al.
\newblock {Feebly-interacting particles: FIPs 2020 workshop report}.
\newblock \emph{Eur. Phys. J. C}, 81\penalty0 (11):\penalty0 1015, 2021.
\newblock \doi{10.1140/epjc/s10052-021-09703-7}.

\bibitem[Polchinski(2007)]{Polchinski:1998rr}
J.~Polchinski.
\newblock \emph{{String theory. Vol. 2: Superstring theory and beyond}}.
\newblock Cambridge Monographs on Mathematical Physics. Cambridge University
  Press, 12 2007.
\newblock ISBN 978-0-511-25228-0, 978-0-521-63304-8, 978-0-521-67228-3.
\newblock \doi{10.1017/CBO9780511618123}.

\bibitem[Kim(1987)]{Kim:1986ax}
Jihn~E. Kim.
\newblock {Light Pseudoscalars, Particle Physics and Cosmology}.
\newblock \emph{Phys. Rept.}, 150:\penalty0 1--177, 1987.
\newblock \doi{10.1016/0370-1573(87)90017-2}.

\bibitem[Fayet(1990)]{Fayet:1990wx}
Pierre Fayet.
\newblock {Extra U(1)'s and New Forces}.
\newblock \emph{Nucl. Phys. B}, 347:\penalty0 743--768, 1990.
\newblock \doi{10.1016/0550-3213(90)90381-M}.

\bibitem[Svrcek and Witten(2006)]{Svrcek:2006yi}
Peter Svrcek and Edward Witten.
\newblock {Axions In String Theory}.
\newblock \emph{JHEP}, 06:\penalty0 051, 2006.
\newblock \doi{10.1088/1126-6708/2006/06/051}.

\bibitem[Arvanitaki et~al.(2010)Arvanitaki, Dimopoulos, Dubovsky, Kaloper, and
  March-Russell]{Arvanitaki_2010}
Asimina Arvanitaki, Savas Dimopoulos, Sergei Dubovsky, Nemanja Kaloper, and
  John March-Russell.
\newblock String axiverse.
\newblock \emph{Phys. Rev. D}, 81:\penalty0 123530, Jun 2010.
\newblock \doi{10.1103/PhysRevD.81.123530}.
\newblock URL \url{https://link.aps.org/doi/10.1103/PhysRevD.81.123530}.

\bibitem[Arias et~al.(2012)Arias, Cadamuro, Goodsell, Jaeckel, Redondo, and
  Ringwald]{Arias:2012az}
Paola Arias, Davide Cadamuro, Mark Goodsell, Joerg Jaeckel, Javier Redondo, and
  Andreas Ringwald.
\newblock {WISPy Cold Dark Matter}.
\newblock \emph{JCAP}, 06:\penalty0 013, 2012.
\newblock \doi{10.1088/1475-7516/2012/06/013}.

\bibitem[Lyth and Riotto(1999)]{Lyth:1998xn}
David~H. Lyth and Antonio Riotto.
\newblock {Particle physics models of inflation and the cosmological density
  perturbation}.
\newblock \emph{Phys. Rept.}, 314:\penalty0 1--146, 1999.
\newblock \doi{10.1016/S0370-1573(98)00128-8}.

\bibitem[Graham et~al.(2016)Graham, Mardon, and Rajendran]{PhysRevD.93.103520}
Peter~W. Graham, Jeremy Mardon, and Surjeet Rajendran.
\newblock Vector dark matter from inflationary fluctuations.
\newblock \emph{Phys. Rev. D}, 93:\penalty0 103520, May 2016.
\newblock \doi{10.1103/PhysRevD.93.103520}.
\newblock URL \url{https://link.aps.org/doi/10.1103/PhysRevD.93.103520}.

\bibitem[Ema et~al.(2019)Ema, Nakayama, and Tang]{Ema:2019yrd}
Yohei Ema, Kazunori Nakayama, and Yong Tang.
\newblock {Production of purely gravitational dark matter: the case of fermion
  and vector boson}.
\newblock \emph{JHEP}, 07:\penalty0 060, 2019.
\newblock \doi{10.1007/JHEP07(2019)060}.

\bibitem[Ahmed et~al.(2020)Ahmed, Grzadkowski, and Socha]{Ahmed:2020fhc}
Aqeel Ahmed, Bohdan Grzadkowski, and Anna Socha.
\newblock {Gravitational production of vector dark matter}.
\newblock \emph{JHEP}, 08:\penalty0 059, 2020.
\newblock \doi{10.1007/JHEP08(2020)059}.

\bibitem[Kolb and Long(2021)]{Kolb:2020fwh}
Edward~W. Kolb and Andrew~J. Long.
\newblock {Completely dark photons from gravitational particle production
  during the inflationary era}.
\newblock \emph{JHEP}, 03:\penalty0 283, 2021.
\newblock \doi{10.1007/JHEP03(2021)283}.

\bibitem[Bar et~al.(2018)Bar, Blas, Blum, and Sibiryakov]{Bar:2018acw}
Nitsan Bar, Diego Blas, Kfir Blum, and Sergey Sibiryakov.
\newblock {Galactic rotation curves versus ultralight dark matter: Implications
  of the soliton-host halo relation}.
\newblock \emph{Phys. Rev. D}, 98\penalty0 (8):\penalty0 083027, 2018.
\newblock \doi{10.1103/PhysRevD.98.083027}.

\bibitem[Marsh and Niemeyer(2019)]{Marsh:2018zyw}
David J.~E. Marsh and Jens~C. Niemeyer.
\newblock {Strong Constraints on Fuzzy Dark Matter from Ultrafaint Dwarf Galaxy
  Eridanus II}.
\newblock \emph{Phys. Rev. Lett.}, 123\penalty0 (5):\penalty0 051103, 2019.
\newblock \doi{10.1103/PhysRevLett.123.051103}.

\bibitem[Pozo et~al.(2024{\natexlab{a}})Pozo, Broadhurst, Smoot, Chiueh, Luu,
  Vogelsberger, and Mocz]{Pozo:2023zmx}
Alvaro Pozo, Tom Broadhurst, George~F. Smoot, Tzihong Chiueh, Hoang~Nhan Luu,
  Mark Vogelsberger, and Philip Mocz.
\newblock {Dwarf galaxies united by dark bosons}.
\newblock \emph{Phys. Rev. D}, 109\penalty0 (8):\penalty0 083532,
  2024{\natexlab{a}}.
\newblock \doi{10.1103/PhysRevD.109.083532}.

\bibitem[Pozo et~al.(2024{\natexlab{b}})Pozo, Broadhurst, de~Martino, Chiueh,
  Smoot, Bonoli, and Angulo]{Pozo:2020ukk}
Alvaro Pozo, Tom Broadhurst, Ivan de~Martino, Tzihong Chiueh, George~F. Smoot,
  Silvia Bonoli, and Raul Angulo.
\newblock {Detection of a universal core-halo transition in dwarf galaxies as
  predicted by Bose-Einstein dark matter}.
\newblock \emph{Phys. Rev. D}, 110\penalty0 (4):\penalty0 043534,
  2024{\natexlab{b}}.
\newblock \doi{10.1103/PhysRevD.110.043534}.

\bibitem[Zimmermann et~al.(2025)Zimmermann, Alvey, Marsh, Fairbairn, and
  Read]{PhysRevLett.134.151001}
Tim Zimmermann, James Alvey, David J.~E. Marsh, Malcolm Fairbairn, and
  Justin~I. Read.
\newblock Dwarf galaxies imply dark matter is heavier than
  $2.2\ifmmode\times\else\texttimes\fi{}{10}^{\ensuremath{-}21}\text{ }\text{
  }\mathrm{eV}$.
\newblock \emph{Phys. Rev. Lett.}, 134:\penalty0 151001, Apr 2025.
\newblock \doi{10.1103/PhysRevLett.134.151001}.
\newblock URL \url{https://link.aps.org/doi/10.1103/PhysRevLett.134.151001}.

\bibitem[May et~al.(2025)May, Dalal, and
  Kravtsov]{may2025updatedboundsultralightdark}
Simon May, Neal Dalal, and Andrey Kravtsov.
\newblock Updated bounds on ultra-light dark matter from the tiniest galaxies,
  2025.
\newblock URL \url{https://arxiv.org/abs/2509.02781}.

\bibitem[Church et~al.(2019)Church, Ostriker, and Mocz]{Church:2018sro}
Benjamin~V. Church, Jeremiah~P. Ostriker, and Philip Mocz.
\newblock {Heating of Milky Way disc Stars by Dark Matter Fluctuations in Cold
  Dark Matter and Fuzzy Dark Matter Paradigms}.
\newblock \emph{Mon. Not. Roy. Astron. Soc.}, 485\penalty0 (2):\penalty0
  2861--2876, 2019.
\newblock \doi{10.1093/mnras/stz534}.

\bibitem[Amorisco and Loeb(2018)]{Amorisco:2018dcn}
Nicola~C. Amorisco and A.~Loeb.
\newblock {First constraints on Fuzzy Dark Matter from the dynamics of stellar
  streams in the Milky Way}.
\newblock 8 2018.

\bibitem[Armengaud et~al.(2017)Armengaud, Palanque-Delabrouille, Y{\`e}che,
  Marsh, and Baur]{Armengaud:2017nkf}
Eric Armengaud, Nathalie Palanque-Delabrouille, Christophe Y{\`e}che, David
  J.~E. Marsh, and Julien Baur.
\newblock {Constraining the mass of light bosonic dark matter using SDSS
  Lyman-$\alpha$ forest}.
\newblock \emph{Mon. Not. Roy. Astron. Soc.}, 471\penalty0 (4):\penalty0
  4606--4614, 2017.
\newblock \doi{10.1093/mnras/stx1870}.

\bibitem[Ir\ifmmode \check{s}\else \v{s}\fi{}i\ifmmode~\check{c}\else
  \v{c}\fi{} et~al.(2017)Ir\ifmmode \check{s}\else
  \v{s}\fi{}i\ifmmode~\check{c}\else \v{c}\fi{}, Viel, Haehnelt, Bolton, and
  Becker]{PhysRevLett.119.031302}
Vid Ir\ifmmode \check{s}\else \v{s}\fi{}i\ifmmode~\check{c}\else \v{c}\fi{},
  Matteo Viel, Martin~G. Haehnelt, James~S. Bolton, and George~D. Becker.
\newblock First constraints on fuzzy dark matter from
  lyman-$\ensuremath{\alpha}$ forest data and hydrodynamical simulations.
\newblock \emph{Phys. Rev. Lett.}, 119:\penalty0 031302, Jul 2017.
\newblock \doi{10.1103/PhysRevLett.119.031302}.
\newblock URL \url{https://link.aps.org/doi/10.1103/PhysRevLett.119.031302}.

\bibitem[Zhang et~al.(2018)Zhang, Kuo, Liu, Tsai, Cheung, and
  Chu]{Zhang:2017chj}
Jiajun Zhang, Jui-Lin Kuo, Hantao Liu, Yue-Lin~Sming Tsai, Kingman Cheung, and
  Ming-Chung Chu.
\newblock {The Importance of Quantum Pressure of Fuzzy Dark Matter on
  Lyman-Alpha Forest}.
\newblock \emph{Astrophys. J.}, 863:\penalty0 73, 2018.
\newblock \doi{10.3847/1538-4357/aacf3f}.

\bibitem[Kobayashi et~al.(2017)Kobayashi, Murgia, De~Simone, Ir{\v{s}}i{\v{c}},
  and Viel]{Kobayashi:2017jcf}
Takeshi Kobayashi, Riccardo Murgia, Andrea De~Simone, Vid Ir{\v{s}}i{\v{c}},
  and Matteo Viel.
\newblock {Lyman-$\alpha$ constraints on ultralight scalar dark matter:
  Implications for the early and late universe}.
\newblock \emph{Phys. Rev. D}, 96\penalty0 (12):\penalty0 123514, 2017.
\newblock \doi{10.1103/PhysRevD.96.123514}.

\bibitem[Schneider(2018)]{Schneider:2018xba}
Aurel Schneider.
\newblock {Constraining noncold dark matter models with the global 21-cm
  signal}.
\newblock \emph{Phys. Rev. D}, 98\penalty0 (6):\penalty0 063021, 2018.
\newblock \doi{10.1103/PhysRevD.98.063021}.

\bibitem[Lidz and Hui(2018)]{Lidz:2018fqo}
Adam Lidz and Lam Hui.
\newblock {Implications of a prereionization 21-cm absorption signal for fuzzy
  dark matter}.
\newblock \emph{Phys. Rev. D}, 98\penalty0 (2):\penalty0 023011, 2018.
\newblock \doi{10.1103/PhysRevD.98.023011}.

\bibitem[Khmelnitsky and Rubakov(2014)]{Khmelnitsky:2013lxt}
Andrei Khmelnitsky and Valery Rubakov.
\newblock {Pulsar timing signal from ultralight scalar dark matter}.
\newblock \emph{JCAP}, 02:\penalty0 019, 2014.
\newblock \doi{10.1088/1475-7516/2014/02/019}.

\bibitem[Porayko and Postnov(2014)]{Porayko:2014rfa}
N.~K. Porayko and K.~A. Postnov.
\newblock {Constraints on ultralight scalar dark matter from pulsar timing}.
\newblock \emph{Phys. Rev. D}, 90\penalty0 (6):\penalty0 062008, 2014.
\newblock \doi{10.1103/PhysRevD.90.062008}.

\bibitem[Nomura et~al.(2020)Nomura, Ito, and Soda]{Nomura:2019cvc}
Kimihiro Nomura, Asuka Ito, and Jiro Soda.
\newblock {Pulsar timing residual induced by ultralight vector dark matter}.
\newblock \emph{Eur. Phys. J. C}, 80\penalty0 (5):\penalty0 419, 2020.
\newblock \doi{10.1140/epjc/s10052-020-7990-y}.

\bibitem[Sun et~al.(2022)Sun, Yang, and Zhang]{Sun_2022}
Sichun Sun, Xing-Yu Yang, and Yun-Long Zhang.
\newblock Pulsar timing residual induced by wideband ultralight dark matter
  with spin 0,1,2.
\newblock \emph{Physical Review D}, 106\penalty0 (6), September 2022.
\newblock ISSN 2470-0029.
\newblock \doi{10.1103/physrevd.106.066006}.
\newblock URL \url{http://dx.doi.org/10.1103/PhysRevD.106.066006}.

\bibitem[Dror and Wei(2025)]{Dror:2025nvg}
Jeff~A. Dror and Qiushi Wei.
\newblock {On Pulsar Timing Detection of Ultralight Vector Dark Matter}.
\newblock 5 2025.

\bibitem[Aoki and Soda(2016)]{Aoki:2016kwl}
Arata Aoki and Jiro Soda.
\newblock {Detecting ultralight axion dark matter wind with laser
  interferometers}.
\newblock \emph{Int. J. Mod. Phys. D}, 26\penalty0 (07):\penalty0 1750063,
  2016.
\newblock \doi{10.1142/S0218271817500638}.

\bibitem[Kim(2023)]{Kim:2023pkx}
Hyungjin Kim.
\newblock {Gravitational interaction of ultralight dark matter with
  interferometers}.
\newblock \emph{JCAP}, 12:\penalty0 018, 2023.
\newblock \doi{10.1088/1475-7516/2023/12/018}.

\bibitem[Yu et~al.(2024)Yu, Cao, Tang, and Wu]{PhysRevD.110.023025}
Jiang-Chuan Yu, Yan Cao, Yong Tang, and Yue-Liang Wu.
\newblock Detecting ultralight dark matter gravitationally with laser
  interferometers in space.
\newblock \emph{Phys. Rev. D}, 110:\penalty0 023025, Jul 2024.
\newblock \doi{10.1103/PhysRevD.110.023025}.
\newblock URL \url{https://link.aps.org/doi/10.1103/PhysRevD.110.023025}.

\bibitem[Smarra et~al.(2023)]{EuropeanPulsarTimingArray:2023egv}
Clemente Smarra et~al.
\newblock {Second Data Release from the European Pulsar Timing Array:
  Challenging the Ultralight Dark Matter Paradigm}.
\newblock \emph{Phys. Rev. Lett.}, 131\penalty0 (17):\penalty0 171001, 2023.
\newblock \doi{10.1103/PhysRevLett.131.171001}.

\bibitem[Gondolo and Silk(1999)]{Gondolo_1999}
Paolo Gondolo and Joseph Silk.
\newblock Dark matter annihilation at the galactic center.
\newblock \emph{Physical Review Letters}, 83\penalty0 (9):\penalty0
  1719–1722, August 1999.
\newblock ISSN 1079-7114.
\newblock \doi{10.1103/physrevlett.83.1719}.
\newblock URL \url{http://dx.doi.org/10.1103/PhysRevLett.83.1719}.

\bibitem[Sadeghian et~al.(2013)Sadeghian, Ferrer, and Will]{Sadeghian:2013laa}
Laleh Sadeghian, Francesc Ferrer, and Clifford~M. Will.
\newblock {Dark matter distributions around massive black holes: A general
  relativistic analysis}.
\newblock \emph{Phys. Rev. D}, 88\penalty0 (6):\penalty0 063522, 2013.
\newblock \doi{10.1103/PhysRevD.88.063522}.

\bibitem[Hu et~al.(2023)Hu, Shao, and Zhang]{Hu:2023ubk}
Zexin Hu, Lijing Shao, and Fupeng Zhang.
\newblock {Prospects for probing small-scale dark matter models with pulsars
  around Sagittarius A*}.
\newblock \emph{Phys. Rev. D}, 108\penalty0 (12):\penalty0 123034, 2023.
\newblock \doi{10.1103/PhysRevD.108.123034}.

\bibitem[Cheng et~al.(2025)Cheng, Cao, and Tang]{Cheng:2024mgl}
Ya-Ze Cheng, Yan Cao, and Yong Tang.
\newblock {Effects of black hole environments on extreme mass-ratio hyperbolic
  encounters}.
\newblock \emph{Phys. Rev. D}, 111\penalty0 (8):\penalty0 083010, 2025.
\newblock \doi{10.1103/PhysRevD.111.083010}.

\bibitem[Amorim et~al.(2019)]{GRAVITY:2019tuf}
A.~Amorim et~al.
\newblock {Scalar field effects on the orbit of S2 star}.
\newblock \emph{Mon. Not. Roy. Astron. Soc.}, 489\penalty0 (4):\penalty0
  4606--4621, 2019.
\newblock \doi{10.1093/mnras/stz2300}.

\bibitem[Yuan et~al.(2022)Yuan, Shen, Tsai, Yuan, and Fan]{Yuan:2022nmu}
Guan-Wen Yuan, Zhao-Qiang Shen, Yue-Lin~Sming Tsai, Qiang Yuan, and Yi-Zhong
  Fan.
\newblock {Constraining ultralight bosonic dark matter with Keck observations
  of S2\textquoteright{}s orbit and kinematics}.
\newblock \emph{Phys. Rev. D}, 106\penalty0 (10):\penalty0 103024, 2022.
\newblock \doi{10.1103/PhysRevD.106.103024}.

\bibitem[Foschi et~al.(2023)]{GRAVITY:2023cjt}
A.~Foschi et~al.
\newblock {Using the motion of S2 to constrain scalar clouds around Sgr A*}.
\newblock \emph{Mon. Not. Roy. Astron. Soc.}, 524\penalty0 (1):\penalty0
  1075--1086, 2023.
\newblock \doi{10.1093/mnras/stad1939}.

\bibitem[Foschi et~al.(2024)]{GRAVITY:2023azi}
A.~Foschi et~al.
\newblock {Using the motion of S2 to constrain vector clouds around Sgr~A*}.
\newblock \emph{Mon. Not. Roy. Astron. Soc.}, 530\penalty0 (4):\penalty0
  3740--3751, 2024.
\newblock \doi{10.1093/mnras/stae423}.

\bibitem[Bar et~al.(2019)Bar, Blum, Lacroix, and Panci]{Bar:2019pnz}
Nitsan Bar, Kfir Blum, Thomas Lacroix, and Paolo Panci.
\newblock {Looking for ultralight dark matter near supermassive black holes}.
\newblock \emph{JCAP}, 07:\penalty0 045, 2019.
\newblock \doi{10.1088/1475-7516/2019/07/045}.

\bibitem[Lacroix(2018)]{Lacroix:2018zmg}
Thomas Lacroix.
\newblock {Dynamical constraints on a dark matter spike at the Galactic Centre
  from stellar orbits}.
\newblock \emph{Astron. Astrophys.}, 619:\penalty0 A46, 2018.
\newblock \doi{10.1051/0004-6361/201832652}.

\bibitem[Chan and Lee(2022)]{Chan_2022}
Man~Ho Chan and Chak~Man Lee.
\newblock Crossing the dark matter soliton core: A possible reversed orbital
  precession.
\newblock \emph{Physical Review D}, 106\penalty0 (12), December 2022.
\newblock ISSN 2470-0029.
\newblock \doi{10.1103/physrevd.106.123018}.
\newblock URL \url{http://dx.doi.org/10.1103/PhysRevD.106.123018}.

\bibitem[Shen et~al.(2023)Shen, Yuan, Jiang, Tsai, Yuan, and Fan]{Shen:2023kkm}
Zhao-Qiang Shen, Guan-Wen Yuan, Cheng-Zi Jiang, Yue-Lin~Sming Tsai, Qiang Yuan,
  and Yi-Zhong Fan.
\newblock {Exploring dark matter spike distribution around the Galactic centre
  with stellar orbits}.
\newblock \emph{Mon. Not. Roy. Astron. Soc.}, 527\penalty0 (2):\penalty0
  3196--3207, 2023.
\newblock \doi{10.1093/mnras/stad3282}.

\bibitem[Zakharov et~al.(2007)Zakharov, Nucita, De~Paolis, and
  Ingrosso]{Zakharov:2007fj}
A.~F. Zakharov, A.~A. Nucita, F.~De~Paolis, and G.~Ingrosso.
\newblock {Apoastron Shift Constraints on Dark Matter Distribution at the
  Galactic Center}.
\newblock \emph{Phys. Rev. D}, 76:\penalty0 062001, 2007.
\newblock \doi{10.1103/PhysRevD.76.062001}.

\bibitem[Hei{\ss}el et~al.(2022)Hei{\ss}el, Paumard, Perrin, and
  Vincent]{Heissel:2021pcw}
Gernot Hei{\ss}el, Thibaut Paumard, Guy Perrin, and Fr{\'e}d{\'e}ric Vincent.
\newblock {The dark mass signature in the orbit of S2}.
\newblock \emph{Astron. Astrophys.}, 660:\penalty0 A13, 2022.
\newblock \doi{10.1051/0004-6361/202142114}.

\bibitem[Detweiler(1980)]{PhysRevD.22.2323}
Steven Detweiler.
\newblock Klein-gordon equation and rotating black holes.
\newblock \emph{Phys. Rev. D}, 22:\penalty0 2323--2326, Nov 1980.
\newblock \doi{10.1103/PhysRevD.22.2323}.
\newblock URL \url{https://link.aps.org/doi/10.1103/PhysRevD.22.2323}.

\bibitem[Cardoso and Yoshida(2005)]{Cardoso:2005vk}
Vitor Cardoso and Shijun Yoshida.
\newblock {Superradiant instabilities of rotating black branes and strings}.
\newblock \emph{JHEP}, 07:\penalty0 009, 2005.
\newblock \doi{10.1088/1126-6708/2005/07/009}.

\bibitem[Dolan(2007{\natexlab{a}})]{PhysRevD.76.084001}
Sam~R. Dolan.
\newblock Instability of the massive klein-gordon field on the kerr spacetime.
\newblock \emph{Phys. Rev. D}, 76:\penalty0 084001, Oct 2007{\natexlab{a}}.
\newblock \doi{10.1103/PhysRevD.76.084001}.
\newblock URL \url{https://link.aps.org/doi/10.1103/PhysRevD.76.084001}.

\bibitem[Arvanitaki and Dubovsky(2011)]{Arvanitaki:2010sy}
Asimina Arvanitaki and Sergei Dubovsky.
\newblock {Exploring the String Axiverse with Precision Black Hole Physics}.
\newblock \emph{Phys. Rev. D}, 83:\penalty0 044026, 2011.
\newblock \doi{10.1103/PhysRevD.83.044026}.

\bibitem[Witek et~al.(2013)Witek, Cardoso, Ishibashi, and
  Sperhake]{PhysRevD.87.043513}
Helvi Witek, Vitor Cardoso, Akihiro Ishibashi, and Ulrich Sperhake.
\newblock Superradiant instabilities in astrophysical systems.
\newblock \emph{Phys. Rev. D}, 87:\penalty0 043513, Feb 2013.
\newblock \doi{10.1103/PhysRevD.87.043513}.
\newblock URL \url{https://link.aps.org/doi/10.1103/PhysRevD.87.043513}.

\bibitem[Brito et~al.(2015{\natexlab{a}})Brito, Cardoso, and
  Pani]{Brito:2014wla}
Richard Brito, Vitor Cardoso, and Paolo Pani.
\newblock {Black holes as particle detectors: evolution of superradiant
  instabilities}.
\newblock \emph{Class. Quant. Grav.}, 32\penalty0 (13):\penalty0 134001,
  2015{\natexlab{a}}.
\newblock \doi{10.1088/0264-9381/32/13/134001}.

\bibitem[Brito et~al.(2015{\natexlab{b}})Brito, Cardoso, and
  Pani]{Brito:2015oca}
Richard Brito, Vitor Cardoso, and Paolo Pani.
\newblock {Superradiance}: {New Frontiers in Black Hole Physics}.
\newblock \emph{Lect. Notes Phys.}, 906:\penalty0 pp.1--237,
  2015{\natexlab{b}}.
\newblock \doi{10.1007/978-3-319-19000-6}.

\bibitem[Baryakhtar et~al.(2017)Baryakhtar, Lasenby, and
  Teo]{Baryakhtar:2017ngi}
Masha Baryakhtar, Robert Lasenby, and Mae Teo.
\newblock {Black Hole Superradiance Signatures of Ultralight Vectors}.
\newblock \emph{Phys. Rev. D}, 96\penalty0 (3):\penalty0 035019, 2017.
\newblock \doi{10.1103/PhysRevD.96.035019}.

\bibitem[Frolov et~al.(2018)Frolov, Krtou\v{s}, Kubiz\v{n}\'ak, and
  Santos]{Frolov:2018ezx}
Valeri~P. Frolov, Pavel Krtou\v{s}, David Kubiz\v{n}\'ak, and Jorge~E. Santos.
\newblock {Massive Vector Fields in Rotating Black-Hole Spacetimes:
  Separability and Quasinormal Modes}.
\newblock \emph{Phys. Rev. Lett.}, 120:\penalty0 231103, 2018.
\newblock \doi{10.1103/PhysRevLett.120.231103}.

\bibitem[Dolan(2018{\natexlab{a}})]{Dolan:2018dqv}
Sam~R. Dolan.
\newblock {Instability of the Proca field on Kerr spacetime}.
\newblock \emph{Phys. Rev. D}, 98\penalty0 (10):\penalty0 104006,
  2018{\natexlab{a}}.
\newblock \doi{10.1103/PhysRevD.98.104006}.

\bibitem[Brito et~al.(2020)Brito, Grillo, and Pani]{Brito_2020}
Richard Brito, Sara Grillo, and Paolo Pani.
\newblock Black hole superradiant instability from ultralight spin-2 fields.
\newblock \emph{Phys. Rev. Lett.}, 124:\penalty0 211101, May 2020.
\newblock \doi{10.1103/PhysRevLett.124.211101}.
\newblock URL \url{https://link.aps.org/doi/10.1103/PhysRevLett.124.211101}.

\bibitem[Baumann et~al.(2019{\natexlab{a}})Baumann, Chia, Stout, and ter
  Haar]{B2}
Daniel Baumann, Horng~Sheng Chia, John Stout, and Lotte ter Haar.
\newblock {The Spectra of Gravitational Atoms}.
\newblock \emph{JCAP}, 12:\penalty0 006, 2019{\natexlab{a}}.
\newblock \doi{10.1088/1475-7516/2019/12/006}.

\bibitem[Baumann et~al.(2020)Baumann, Chia, Porto, and Stout]{B3}
Daniel Baumann, Horng~Sheng Chia, Rafael~A. Porto, and John Stout.
\newblock {Gravitational Collider Physics}.
\newblock \emph{Phys. Rev. D}, 101\penalty0 (8):\penalty0 083019, 2020.
\newblock \doi{10.1103/PhysRevD.101.083019}.

\bibitem[Siemonsen et~al.(2023)Siemonsen, May, and East]{Siemonsen:2022yyf}
Nils Siemonsen, Taillte May, and William~E. East.
\newblock {Modeling the black hole superradiance gravitational waveform}.
\newblock \emph{Phys. Rev. D}, 107\penalty0 (10):\penalty0 104003, 2023.
\newblock \doi{10.1103/PhysRevD.107.104003}.

\bibitem[Schive et~al.(2014{\natexlab{a}})Schive, Chiueh, and
  Broadhurst]{Schive:2014dra}
Hsi-Yu Schive, Tzihong Chiueh, and Tom Broadhurst.
\newblock {Cosmic Structure as the Quantum Interference of a Coherent Dark
  Wave}.
\newblock \emph{Nature Phys.}, 10:\penalty0 496--499, 2014{\natexlab{a}}.
\newblock \doi{10.1038/nphys2996}.

\bibitem[Schive et~al.(2014{\natexlab{b}})Schive, Liao, Woo, Wong, Chiueh,
  Broadhurst, and Hwang]{Schive:2014hza}
Hsi-Yu Schive, Ming-Hsuan Liao, Tak-Pong Woo, Shing-Kwong Wong, Tzihong Chiueh,
  Tom Broadhurst, and W.~Y.~Pauchy Hwang.
\newblock {Understanding the Core-Halo Relation of Quantum Wave Dark Matter
  from 3D Simulations}.
\newblock \emph{Phys. Rev. Lett.}, 113\penalty0 (26):\penalty0 261302,
  2014{\natexlab{b}}.
\newblock \doi{10.1103/PhysRevLett.113.261302}.

\bibitem[Chavanis(2019)]{Chavanis:2019bnu}
Pierre-Henri Chavanis.
\newblock {Mass-radius relation of self-gravitating Bose-Einstein condensates
  with a central black hole}.
\newblock \emph{Eur. Phys. J. Plus}, 134\penalty0 (7):\penalty0 352, 2019.
\newblock \doi{10.1140/epjp/i2019-12734-7}.

\bibitem[Davies and Mocz(2020)]{Davies_2020}
Elliot~Y Davies and Philip Mocz.
\newblock Fuzzy dark matter soliton cores around supermassive black holes.
\newblock \emph{Monthly Notices of the Royal Astronomical Society},
  492\penalty0 (4):\penalty0 5721–5729, January 2020.
\newblock ISSN 1365-2966.
\newblock \doi{10.1093/mnras/staa202}.
\newblock URL \url{http://dx.doi.org/10.1093/mnras/staa202}.

\bibitem[Annulli et~al.(2020)Annulli, Cardoso, and Vicente]{Annulli_2020}
Lorenzo Annulli, Vitor Cardoso, and Rodrigo Vicente.
\newblock Response of ultralight dark matter to supermassive black holes and
  binaries.
\newblock \emph{Physical Review D}, 102\penalty0 (6), September 2020.
\newblock ISSN 2470-0029.
\newblock \doi{10.1103/physrevd.102.063022}.
\newblock URL \url{http://dx.doi.org/10.1103/PhysRevD.102.063022}.

\bibitem[Zagorac et~al.(2023)Zagorac, Kendall, Padmanabhan, and
  Easther]{Zagorac_2023}
J.~Luna Zagorac, Emily Kendall, Nikhil Padmanabhan, and Richard Easther.
\newblock Soliton formation and the core-halo mass relation: An eigenstate
  perspective.
\newblock \emph{Physical Review D}, 107\penalty0 (8), April 2023.
\newblock ISSN 2470-0029.
\newblock \doi{10.1103/physrevd.107.083513}.
\newblock URL \url{http://dx.doi.org/10.1103/PhysRevD.107.083513}.

\bibitem[Aghaie et~al.(2024)Aghaie, Armando, Dondarini, and Panci]{Aghaie_2024}
Mohammad Aghaie, Giovanni Armando, Alessandro Dondarini, and Paolo Panci.
\newblock Bounds on ultralight dark matter from nanograv.
\newblock \emph{Physical Review D}, 109\penalty0 (10), May 2024.
\newblock ISSN 2470-0029.
\newblock \doi{10.1103/physrevd.109.103030}.
\newblock URL \url{http://dx.doi.org/10.1103/PhysRevD.109.103030}.

\bibitem[Liao et~al.(2025)Liao, Su, Schive, Kunkel, Huang, and
  Chiueh]{liao2025decipheringsolitonhalorelationfuzzy}
Pin-Yu Liao, Guan-Ming Su, Hsi-Yu Schive, Alexander Kunkel, Hsinhao Huang, and
  Tzihong Chiueh.
\newblock Deciphering the soliton-halo relation in fuzzy dark matter, 2025.
\newblock URL \url{https://arxiv.org/abs/2412.09908}.

\bibitem[Chen et~al.(2023)Chen, Xue, Brito, and Cardoso]{Chen_2023}
Yifan Chen, Xiao Xue, Richard Brito, and Vitor Cardoso.
\newblock Photon ring astrometry for superradiant clouds.
\newblock \emph{Physical Review Letters}, 130\penalty0 (11), March 2023.
\newblock ISSN 1079-7114.
\newblock \doi{10.1103/physrevlett.130.111401}.
\newblock URL \url{http://dx.doi.org/10.1103/PhysRevLett.130.111401}.

\bibitem[Chen et~al.(2025)Chen, Xue, and Cardoso]{Chen_2025}
Yifan Chen, Xiao Xue, and Vitor Cardoso.
\newblock Black holes as fermion factories.
\newblock \emph{Journal of Cosmology and Astroparticle Physics}, 2025\penalty0
  (02):\penalty0 035, February 2025.
\newblock ISSN 1475-7516.
\newblock \doi{10.1088/1475-7516/2025/02/035}.
\newblock URL \url{http://dx.doi.org/10.1088/1475-7516/2025/02/035}.

\bibitem[Bai et~al.(2025)Bai, Cardoso, Chen, Do, Hees, Xiao, and
  Xue]{bai2025probingaxionsspectroscopicmeasurements}
Zhaoyu Bai, Vitor Cardoso, Yifan Chen, Tuan Do, Aurélien Hees, Huangyu Xiao,
  and Xiao Xue.
\newblock Probing axions via spectroscopic measurements of s-stars at the
  galactic center, 2025.
\newblock URL \url{https://arxiv.org/abs/2507.07482}.

\bibitem[Tomaselli and Caputo(2025)]{tomaselli2025probingdenseenvironmentssgr}
Giovanni~Maria Tomaselli and Andrea Caputo.
\newblock Probing dense environments around sgr a* with s-stars dynamics, 2025.
\newblock URL \url{https://arxiv.org/abs/2509.03568}.

\bibitem[Wex and Kopeikin(1999)]{Wex:1998wt}
N.~Wex and S.~Kopeikin.
\newblock {Frame dragging and other precessional effects in black hole-pulsar
  binaries}.
\newblock \emph{Astrophys. J.}, 514:\penalty0 388, 1999.
\newblock \doi{10.1086/306933}.

\bibitem[Liu et~al.(2012{\natexlab{a}})Liu, Wex, Kramer, Cordes, and
  Lazio]{Liu:2011ae}
K.~Liu, N.~Wex, M.~Kramer, J.~M. Cordes, and T.~J.~W. Lazio.
\newblock {Prospects for Probing the Spacetime of Sgr A* with Pulsars}.
\newblock \emph{Astrophys. J.}, 747:\penalty0 1, 2012{\natexlab{a}}.
\newblock \doi{10.1088/0004-637X/747/1/1}.

\bibitem[Psaltis et~al.(2016)Psaltis, Wex, and Kramer]{Psaltis:2015uza}
Dimitrios Psaltis, Norbert Wex, and Michael Kramer.
\newblock {A Quantitative Test of the No-Hair Theorem with Sgr A* using stars,
  pulsars, and the Event Horizon Telescope}.
\newblock \emph{Astrophys. J.}, 818\penalty0 (2):\penalty0 121, 2016.
\newblock \doi{10.3847/0004-637X/818/2/121}.

\bibitem[Zhang and Saha(2017)]{Zhang:2017qbb}
Fupeng Zhang and Prasenjit Saha.
\newblock {Probing the spinning of the massive black hole in the Galactic
  Center via pulsar timing: A Full Relativistic Treatment}.
\newblock \emph{Astrophys. J.}, 849\penalty0 (1):\penalty0 33, 2017.
\newblock \doi{10.3847/1538-4357/aa8f47}.

\bibitem[Bower et~al.(2018)]{Bower:2018mta}
Geoffrey~C. Bower et~al.
\newblock {Galactic Center Pulsars with the ngVLA}.
\newblock \emph{ASP Conf. Ser.}, 517:\penalty0 793, 2018.

\bibitem[Dong et~al.(2022)Dong, Shao, Hu, Miao, and Wang]{Dong:2022zvh}
Yiming Dong, Lijing Shao, Zexin Hu, Xueli Miao, and Ziming Wang.
\newblock {Prospects for constraining the Yukawa gravity with pulsars around
  Sagittarius~A*}.
\newblock \emph{JCAP}, 11:\penalty0 051, 2022.
\newblock \doi{10.1088/1475-7516/2022/11/051}.

\bibitem[Hu et~al.(2024)Hu, Shao, Xu, Liang, and Mai]{Hu:2023vsg}
Zexin Hu, Lijing Shao, Rui Xu, Dicong Liang, and Zhan-Feng Mai.
\newblock {Probing the vector charge of Sagittarius A* with pulsar timing}.
\newblock \emph{JCAP}, 04:\penalty0 087, 2024.
\newblock \doi{10.1088/1475-7516/2024/04/087}.

\bibitem[Hu and Shao(2024)]{Hu:2024blq}
Zexin Hu and Lijing Shao.
\newblock {Measuring the Spin of the Galactic Center Supermassive Black Hole
  with Two Pulsars}.
\newblock \emph{Phys. Rev. Lett.}, 133\penalty0 (23):\penalty0 231402, 2024.
\newblock \doi{10.1103/PhysRevLett.133.231402}.

\bibitem[Shao and Hu(2025)]{Shao:2025vmb}
Lijing Shao and Zexin Hu.
\newblock {Fundamental Physics with Pulsars around Sagittarius A$^\star$}.
\newblock 2025.
\newblock \doi{10.48550/arXiv.2508.09931}.

\bibitem[De~Martino et~al.(2017)De~Martino, Broadhurst, Henry~Tye, Chiueh,
  Schive, and Lazkoz]{DeMartino:2017qsa}
Ivan De~Martino, Tom Broadhurst, S.~H. Henry~Tye, Tzihong Chiueh, Hsi-Yu
  Schive, and Ruth Lazkoz.
\newblock {Recognizing Axionic Dark Matter by Compton and de Broglie Scale
  Modulation of Pulsar Timing}.
\newblock \emph{Phys. Rev. Lett.}, 119\penalty0 (22):\penalty0 221103, 2017.
\newblock \doi{10.1103/PhysRevLett.119.221103}.

\bibitem[Jain and Amin(2022)]{Jain:2021pnk}
Mudit Jain and Mustafa~A. Amin.
\newblock {Polarized solitons in higher-spin wave dark matter}.
\newblock \emph{Phys. Rev. D}, 105\penalty0 (5):\penalty0 056019, 2022.
\newblock \doi{10.1103/PhysRevD.105.056019}.

\bibitem[Cao and Tang(2023)]{Cao:2023fyv}
Yan Cao and Yong Tang.
\newblock {Signatures of ultralight bosons in compact binary inspiral and
  outspiral}.
\newblock \emph{Phys. Rev. D}, 108\penalty0 (12):\penalty0 123017, 2023.
\newblock \doi{10.1103/PhysRevD.108.123017}.

\bibitem[Cao et~al.(2025)Cao, Cheng, Li, and Tang]{Cao:2024wby}
Yan Cao, Ya-Ze Cheng, Gen-Liang Li, and Yong Tang.
\newblock {Probing vector gravitational atoms with eccentric intermediate
  mass-ratio inspirals}.
\newblock \emph{Phys. Rev. D}, 111\penalty0 (8):\penalty0 083011, 2025.
\newblock \doi{10.1103/PhysRevD.111.083011}.

\bibitem[Sup()]{SupplementaryMaterials}
Supplementary materials.

\bibitem[Salehian et~al.(2021)Salehian, Zhang, Amin, Kaiser, and
  Namjoo]{Salehian_2021}
Borna Salehian, Hong-Yi Zhang, Mustafa~A. Amin, David~I. Kaiser, and
  Mohammad~Hossein Namjoo.
\newblock Beyond schrödinger-poisson: nonrelativistic effective field theory
  for scalar dark matter.
\newblock \emph{Journal of High Energy Physics}, 2021\penalty0 (9), September
  2021.
\newblock ISSN 1029-8479.
\newblock \doi{10.1007/jhep09(2021)050}.
\newblock URL \url{http://dx.doi.org/10.1007/JHEP09(2021)050}.

\bibitem[Zhang(2025)]{zhang2025unifiedviewscalarvector}
Hong-Yi Zhang.
\newblock Unified view of scalar and vector dark matter solitons, 2025.
\newblock URL \url{https://arxiv.org/abs/2406.05031}.

\bibitem[Beloborodov et~al.(2006)Beloborodov, Levin, Eisenhauer, Genzel,
  Paumard, Gillessen, and Ott]{Beloborodov:2006is}
Andrei~M. Beloborodov, Yuri Levin, Frank Eisenhauer, Reinhard Genzel, Thibaut
  Paumard, Stefan Gillessen, and Thomas Ott.
\newblock {Clockwise stellar disk and the dark mass in the galactic center}.
\newblock \emph{Astrophys. J.}, 648:\penalty0 405--410, 2006.
\newblock \doi{10.1086/504279}.

\bibitem[Abuter et~al.(2020{\natexlab{a}})]{GRAVITY:2020gka}
R.~Abuter et~al.
\newblock {Detection of the Schwarzschild precession in the orbit of the star
  S2 near the Galactic centre massive black hole}.
\newblock \emph{Astron. Astrophys.}, 636:\penalty0 L5, 2020{\natexlab{a}}.
\newblock \doi{10.1051/0004-6361/202037813}.

\bibitem[Ficarra et~al.(2019)Ficarra, Pani, and Witek]{Ficarra_2019}
Giuseppe Ficarra, Paolo Pani, and Helvi Witek.
\newblock Impact of multiple modes on the black-hole superradiant instability.
\newblock \emph{Physical Review D}, 99\penalty0 (10), May 2019.
\newblock ISSN 2470-0029.
\newblock \doi{10.1103/physrevd.99.104019}.
\newblock URL \url{http://dx.doi.org/10.1103/PhysRevD.99.104019}.

\bibitem[Siemonsen and East(2020)]{Siemonsen_2020}
Nils Siemonsen and William~E. East.
\newblock Gravitational wave signatures of ultralight vector bosons from black
  hole superradiance.
\newblock \emph{Physical Review D}, 101\penalty0 (2), January 2020.
\newblock ISSN 2470-0029.
\newblock \doi{10.1103/physrevd.101.024019}.
\newblock URL \url{http://dx.doi.org/10.1103/PhysRevD.101.024019}.

\bibitem[Guo et~al.(2023)Guo, Bao, and Zhang]{Guo_2023}
Yin-da Guo, Shou-shan Bao, and Hong Zhang.
\newblock Subdominant modes of the scalar superradiant instability and
  gravitational wave beats.
\newblock \emph{Physical Review D}, 107\penalty0 (7), April 2023.
\newblock ISSN 2470-0029.
\newblock \doi{10.1103/physrevd.107.075009}.
\newblock URL \url{http://dx.doi.org/10.1103/PhysRevD.107.075009}.

\bibitem[Guo et~al.(2025{\natexlab{a}})Guo, Bao, Li, and
  Zhang]{guo2025effectaccretionscalarsuperradiant}
Yin-Da Guo, Shou-Shan Bao, Tianjun Li, and Hong Zhang.
\newblock The effect of accretion on scalar superradiant instability,
  2025{\natexlab{a}}.
\newblock URL \url{https://arxiv.org/abs/2501.09280}.

\bibitem[Zagorac et~al.(2022)Zagorac, Sands, Padmanabhan, and
  Easther]{Zagorac_2022}
J.~Luna Zagorac, Isabel Sands, Nikhil Padmanabhan, and Richard Easther.
\newblock Schrödinger-poisson solitons: Perturbation theory.
\newblock \emph{Physical Review D}, 105\penalty0 (10), May 2022.
\newblock ISSN 2470-0029.
\newblock \doi{10.1103/physrevd.105.103506}.
\newblock URL \url{http://dx.doi.org/10.1103/PhysRevD.105.103506}.

\bibitem[Glennon et~al.(2023)Glennon, Mirasola, Musoke, Neyrinck, and
  Prescod-Weinstein]{Glennon_2023}
Noah Glennon, Anthony~E. Mirasola, Nathan Musoke, Mark~C. Neyrinck, and Chanda
  Prescod-Weinstein.
\newblock Scalar dark matter vortex stabilization with black holes.
\newblock \emph{Journal of Cosmology and Astroparticle Physics}, 2023\penalty0
  (07):\penalty0 004, July 2023.
\newblock ISSN 1475-7516.
\newblock \doi{10.1088/1475-7516/2023/07/004}.
\newblock URL \url{http://dx.doi.org/10.1088/1475-7516/2023/07/004}.

\bibitem[Salasnich and
  Yakimenko(2025)]{salasnich2025collectiveexcitationsselfgravitatingultralight}
Luca Salasnich and Alexander Yakimenko.
\newblock Collective excitations of self-gravitating ultralight dark matter
  cores, 2025.
\newblock URL \url{https://arxiv.org/abs/2501.06891}.

\bibitem[Merritt et~al.(2010)Merritt, Alexander, Mikkola, and
  Will]{PhysRevD.81.062002}
David Merritt, Tal Alexander, Seppo Mikkola, and Clifford~M. Will.
\newblock Testing properties of the galactic center black hole using stellar
  orbits.
\newblock \emph{Phys. Rev. D}, 81:\penalty0 062002, Mar 2010.
\newblock \doi{10.1103/PhysRevD.81.062002}.
\newblock URL \url{https://link.aps.org/doi/10.1103/PhysRevD.81.062002}.

\bibitem[Liu et~al.(2012{\natexlab{b}})Liu, Wex, Kramer, Cordes, and
  Lazio]{Liu_2012}
K.~Liu, N.~Wex, M.~Kramer, J.~M. Cordes, and T.~J.~W. Lazio.
\newblock Prospects for probing the spacetime of sgr a* with pulsars.
\newblock \emph{The Astrophysical Journal}, 747\penalty0 (1):\penalty0 1, feb
  2012{\natexlab{b}}.
\newblock \doi{10.1088/0004-637X/747/1/1}.
\newblock URL \url{https://dx.doi.org/10.1088/0004-637X/747/1/1}.

\bibitem[Du et~al.(2022)Du, Egaña-Ugrinovic, Essig, Fragione, and
  Perna]{Du_2022}
Peizhi Du, Daniel Egaña-Ugrinovic, Rouven Essig, Giacomo Fragione, and Rosalba
  Perna.
\newblock Searching for ultra-light bosons and constraining black hole spin
  distributions with stellar tidal disruption events.
\newblock \emph{Nature Communications}, 13\penalty0 (1), August 2022.
\newblock ISSN 2041-1723.
\newblock \doi{10.1038/s41467-022-32301-4}.
\newblock URL \url{http://dx.doi.org/10.1038/s41467-022-32301-4}.

\bibitem[Thorne(1980)]{RevModPhys.52.299}
Kip~S. Thorne.
\newblock Multipole expansions of gravitational radiation.
\newblock \emph{Rev. Mod. Phys.}, 52:\penalty0 299--339, Apr 1980.
\newblock \doi{10.1103/RevModPhys.52.299}.
\newblock URL \url{https://link.aps.org/doi/10.1103/RevModPhys.52.299}.

\bibitem[Ferreira et~al.(2017)Ferreira, Macedo, and Cardoso]{Ferreira:2017pth}
Miguel~C. Ferreira, Caio F.~B. Macedo, and Vitor Cardoso.
\newblock {Orbital fingerprints of ultralight scalar fields around black
  holes}.
\newblock \emph{Phys. Rev. D}, 96\penalty0 (8):\penalty0 083017, 2017.
\newblock \doi{10.1103/PhysRevD.96.083017}.

\bibitem[Baumann et~al.(2019{\natexlab{b}})Baumann, Chia, and Porto]{B1}
Daniel Baumann, Horng~Sheng Chia, and Rafael~A. Porto.
\newblock {Probing Ultralight Bosons with Binary Black Holes}.
\newblock \emph{Phys. Rev. D}, 99\penalty0 (4):\penalty0 044001,
  2019{\natexlab{b}}.
\newblock \doi{10.1103/PhysRevD.99.044001}.

\bibitem[Baumann et~al.(2022)Baumann, Bertone, Stout, and Tomaselli]{B4}
Daniel Baumann, Gianfranco Bertone, John Stout, and Giovanni~Maria Tomaselli.
\newblock Ionization of gravitational atoms.
\newblock \emph{Phys. Rev. D}, 105:\penalty0 115036, Jun 2022.
\newblock \doi{10.1103/PhysRevD.105.115036}.
\newblock URL \url{https://link.aps.org/doi/10.1103/PhysRevD.105.115036}.

\bibitem[Tomaselli et~al.(2023)Tomaselli, Spieksma, and
  Bertone]{Tomaselli:2023ysb}
Giovanni~Maria Tomaselli, Thomas F.~M. Spieksma, and Gianfranco Bertone.
\newblock {Dynamical friction in gravitational atoms}.
\newblock \emph{JCAP}, 07:\penalty0 070, 2023.
\newblock \doi{10.1088/1475-7516/2023/07/070}.

\bibitem[Tomaselli et~al.(2024)Tomaselli, Spieksma, and
  Bertone]{Tomaselli:2024bdd}
Giovanni~Maria Tomaselli, Thomas F.~M. Spieksma, and Gianfranco Bertone.
\newblock Resonant history of gravitational atoms in black hole binaries.
\newblock \emph{Phys. Rev. D}, 110:\penalty0 064048, Sep 2024.
\newblock \doi{10.1103/PhysRevD.110.064048}.
\newblock URL \url{https://link.aps.org/doi/10.1103/PhysRevD.110.064048}.

\bibitem[Bo\ifmmode \check{s}\else \v{s}\fi{}kovi\ifmmode~\acute{c}\else
  \'{c}\fi{} et~al.(2024)Bo\ifmmode \check{s}\else
  \v{s}\fi{}kovi\ifmmode~\acute{c}\else \'{c}\fi{}, Koschnitzke, and
  Porto]{Boskovic:2024fga}
Mateja Bo\ifmmode \check{s}\else \v{s}\fi{}kovi\ifmmode~\acute{c}\else
  \'{c}\fi{}, Matthias Koschnitzke, and Rafael~A. Porto.
\newblock Signatures of ultralight bosons in the orbital eccentricity of binary
  black holes.
\newblock \emph{Phys. Rev. Lett.}, 133:\penalty0 121401, Sep 2024.
\newblock \doi{10.1103/PhysRevLett.133.121401}.
\newblock URL \url{https://link.aps.org/doi/10.1103/PhysRevLett.133.121401}.

\bibitem[Tomaselli(2025)]{Tomaselli:2025jfo}
Giovanni~Maria Tomaselli.
\newblock {Smooth binary evolution from wide resonances in boson clouds}.
\newblock 7 2025.

\bibitem[Brito and Shah(2023)]{Brito_2023}
Richard Brito and Shreya Shah.
\newblock Extreme mass-ratio inspirals into black holes surrounded by scalar
  clouds.
\newblock \emph{Physical Review D}, 108\penalty0 (8), October 2023.
\newblock ISSN 2470-0029.
\newblock \doi{10.1103/physrevd.108.084019}.
\newblock URL \url{http://dx.doi.org/10.1103/PhysRevD.108.084019}.

\bibitem[Duque et~al.(2024)Duque, Macedo, Vicente, and Cardoso]{Duque:2023cac}
Francisco Duque, Caio F.~B. Macedo, Rodrigo Vicente, and Vitor Cardoso.
\newblock Extreme-mass-ratio inspirals in ultralight dark matter.
\newblock \emph{Phys. Rev. Lett.}, 133:\penalty0 121404, Sep 2024.
\newblock \doi{10.1103/PhysRevLett.133.121404}.
\newblock URL \url{https://link.aps.org/doi/10.1103/PhysRevLett.133.121404}.

\bibitem[Dyson et~al.(2025)Dyson, Spieksma, Brito, van~de Meent, and
  Dolan]{dyson2025environmentaleffectsextrememass}
Conor Dyson, Thomas F.~M. Spieksma, Richard Brito, Maarten van~de Meent, and
  Sam Dolan.
\newblock Environmental effects in extreme mass ratio inspirals: perturbations
  to the environment in kerr, 2025.
\newblock URL \url{https://arxiv.org/abs/2501.09806}.

\bibitem[Takahashi et~al.(2022)Takahashi, Omiya, and Tanaka]{Takahashi:2021yhy}
Takuya Takahashi, Hidetoshi Omiya, and Takahiro Tanaka.
\newblock {Axion cloud evaporation during inspiral of black hole binaries: The
  effects of backreaction and radiation}.
\newblock \emph{PTEP}, 2022\penalty0 (4):\penalty0 043E01, 2022.
\newblock \doi{10.1093/ptep/ptac044}.

\bibitem[Takahashi et~al.(2023)Takahashi, Omiya, and Tanaka]{Takahashi:2023flk}
Takuya Takahashi, Hidetoshi Omiya, and Takahiro Tanaka.
\newblock {Evolution of binary systems accompanying axion clouds in extreme
  mass ratio inspirals}.
\newblock \emph{Phys. Rev. D}, 107\penalty0 (10):\penalty0 103020, 2023.
\newblock \doi{10.1103/PhysRevD.107.103020}.

\bibitem[Zhang and Yang(2019)]{Zhang:2018kib}
Jun Zhang and Huan Yang.
\newblock {Gravitational floating orbits around hairy black holes}.
\newblock \emph{Phys. Rev. D}, 99\penalty0 (6):\penalty0 064018, 2019.
\newblock \doi{10.1103/PhysRevD.99.064018}.

\bibitem[Zhang and Yang(2020)]{ZJ}
Jun Zhang and Huan Yang.
\newblock Dynamic signatures of black hole binaries with superradiant clouds.
\newblock \emph{Phys. Rev. D}, 101:\penalty0 043020, Feb 2020.
\newblock \doi{10.1103/PhysRevD.101.043020}.
\newblock URL \url{https://link.aps.org/doi/10.1103/PhysRevD.101.043020}.

\bibitem[Guo et~al.(2024)Guo, Zhang, and Yang]{Guo:2024iye}
Ao~Guo, Jun Zhang, and Huan Yang.
\newblock Superradiant clouds may be relevant for close compact object
  binaries.
\newblock \emph{Phys. Rev. D}, 110:\penalty0 023022, Jul 2024.
\newblock \doi{10.1103/PhysRevD.110.023022}.
\newblock URL \url{https://link.aps.org/doi/10.1103/PhysRevD.110.023022}.

\bibitem[Liu(2024)]{Liu:2024mzw}
Jing Liu.
\newblock Gravitational laser: the stimulated radiation of gravitational waves
  from the clouds of ultralight bosons, 2024.
\newblock URL \url{https://arxiv.org/abs/2401.16096}.

\bibitem[Peng and Zhang(2025)]{Peng:2025zca}
Si-Tong Peng and Jun Zhang.
\newblock {Gravitational Waves from Superradiant Cloud Level Transition}.
\newblock 4 2025.

\bibitem[Guo et~al.(2025{\natexlab{b}})Guo, Zhang, Yang, and
  Zhang]{Guo:2025ckp}
Ao~Guo, Qi-Yan Zhang, Huan Yang, and Jun Zhang.
\newblock {Common Envelope Evolution of Ultralight Boson Clouds}.
\newblock 8 2025{\natexlab{b}}.

\bibitem[Ding et~al.(2021)Ding, Tong, and Wang]{WY_1}
Qianhang Ding, Xi~Tong, and Yi~Wang.
\newblock {Gravitational Collider Physics via Pulsar-Black Hole Binaries}.
\newblock \emph{Astrophys. J.}, 908\penalty0 (1):\penalty0 78, 2021.
\newblock \doi{10.3847/1538-4357/abd803}.

\bibitem[Tong et~al.(2022{\natexlab{a}})Tong, Wang, and Zhu]{WY_2}
Xi~Tong, Yi~Wang, and Hui-Yu Zhu.
\newblock {Gravitational Collider Physics via Pulsar\textendash{}Black Hole
  Binaries II: Fine and Hyperfine Structures Are Favored}.
\newblock \emph{Astrophys. J.}, 924\penalty0 (2):\penalty0 99,
  2022{\natexlab{a}}.
\newblock \doi{10.3847/1538-4357/ac36db}.

\bibitem[Tong et~al.(2022{\natexlab{b}})Tong, Wang, and Zhu]{WY_3}
Xi~Tong, Yi~Wang, and Hui-Yu Zhu.
\newblock {Termination of superradiance from a binary companion}.
\newblock \emph{Phys. Rev. D}, 106\penalty0 (4):\penalty0 043002,
  2022{\natexlab{b}}.
\newblock \doi{10.1103/PhysRevD.106.043002}.

\bibitem[Fan et~al.(2024)Fan, Tong, Wang, and Zhu]{Fan:2023jjj}
Kaiyuan Fan, Xi~Tong, Yi~Wang, and Hui-Yu Zhu.
\newblock {Modulating binary dynamics via the termination of black hole
  superradiance}.
\newblock \emph{Phys. Rev. D}, 109\penalty0 (2):\penalty0 024059, 2024.
\newblock \doi{10.1103/PhysRevD.109.024059}.

\bibitem[Kavic et~al.(2020)Kavic, Liebling, Lippert, and
  Simonetti]{Kavic:2019cgk}
Michael Kavic, Steven~L. Liebling, Matthew Lippert, and John~H. Simonetti.
\newblock {Accessing the axion via compact object binaries}.
\newblock \emph{JCAP}, 08:\penalty0 005, 2020.
\newblock \doi{10.1088/1475-7516/2020/08/005}.

\bibitem[De~Luca and Pani(2021)]{DeLuca:2021ite}
Valerio De~Luca and Paolo Pani.
\newblock {Tidal deformability of dressed black holes and tests of ultralight
  bosons in extended mass ranges}.
\newblock \emph{JCAP}, 08:\penalty0 032, 2021.
\newblock \doi{10.1088/1475-7516/2021/08/032}.

\bibitem[Arana et~al.(2024)Arana, Brito, and Castro]{Arana:2024kaz}
Ricardo Arana, Richard Brito, and Gonçalo Castro.
\newblock Tidal love numbers of gravitational atoms, 2024.

\bibitem[Su et~al.(2021)Su, Xianyu, and Zhang]{Su:2021dwz}
Boye Su, Zhong-Zhi Xianyu, and Xingyu Zhang.
\newblock {Probing Ultralight Bosons with Compact Eccentric Binaries}.
\newblock \emph{Astrophys. J.}, 923\penalty0 (1):\penalty0 114, 2021.
\newblock \doi{10.3847/1538-4357/ac2d91}.

\bibitem[Della~Monica and Brito(2025)]{DellaMonica:2025zby}
Riccardo Della~Monica and Richard Brito.
\newblock {Detectability of gravitational atoms in black hole binaries with the
  Einstein Telescope}.
\newblock \emph{Phys. Rev. D}, 112\penalty0 (2):\penalty0 024074, 2025.
\newblock \doi{10.1103/h7ld-vv9p}.

\bibitem[Lyu et~al.(2025)Lyu, Cai, Guo, He, and Liu]{Lyu:2025lue}
Zhen-Hong Lyu, Rong-Gen Cai, Zong-Kuan Guo, Jian-Feng He, and Jing Liu.
\newblock {Ring formation from black hole superradiance through repeated
  particle production on bound orbits}.
\newblock 7 2025.

\bibitem[Guo et~al.(2025{\natexlab{c}})Guo, Zhong, Chen, Cardoso, Ikeda, and
  Zhou]{Guo:2025pea}
Yuhao Guo, Zhen Zhong, Yifan Chen, Vitor Cardoso, Taishi Ikeda, and Lihang
  Zhou.
\newblock {Ultralight Boson Ionization from Comparable-Mass Binary Black
  Holes}.
\newblock 9 2025{\natexlab{c}}.

\bibitem[Dolan(2007{\natexlab{b}})]{Dolan_2007}
Sam~R. Dolan.
\newblock Instability of the massive klein-gordon field on the kerr spacetime.
\newblock \emph{Physical Review D}, 76\penalty0 (8), October
  2007{\natexlab{b}}.
\newblock ISSN 1550-2368.
\newblock \doi{10.1103/physrevd.76.084001}.
\newblock URL \url{http://dx.doi.org/10.1103/PhysRevD.76.084001}.

\bibitem[Bao et~al.(2022)Bao, Xu, and Zhang]{Bao_2022}
Shou-Shan Bao, Qi-Xuan Xu, and Hong Zhang.
\newblock Improved analytic solution of black hole superradiance.
\newblock \emph{Physical Review D}, 106\penalty0 (6), September 2022.
\newblock ISSN 2470-0029.
\newblock \doi{10.1103/physrevd.106.064016}.
\newblock URL \url{http://dx.doi.org/10.1103/PhysRevD.106.064016}.

\bibitem[Dolan(2018{\natexlab{b}})]{PhysRevD.98.104006}
Sam~R. Dolan.
\newblock Instability of the proca field on kerr spacetime.
\newblock \emph{Phys. Rev. D}, 98:\penalty0 104006, Nov 2018{\natexlab{b}}.
\newblock \doi{10.1103/PhysRevD.98.104006}.
\newblock URL \url{https://link.aps.org/doi/10.1103/PhysRevD.98.104006}.

\bibitem[Fell et~al.(2023)Fell, Heisenberg, and Veske]{Fell_2023}
Shaun Fell, Lavinia Heisenberg, and Doğa Veske.
\newblock Detecting fundamental vector fields with lisa.
\newblock \emph{Physical Review D}, 108\penalty0 (8), October 2023.
\newblock ISSN 2470-0029.
\newblock \doi{10.1103/physrevd.108.083010}.
\newblock URL \url{http://dx.doi.org/10.1103/PhysRevD.108.083010}.

\bibitem[{Damour} and {Deruelle}(1986)]{AIHPA_1986__44_3_263_0}
T.~{Damour} and N.~{Deruelle}.
\newblock {General relativistic celestial mechanics of binary systems. II. The
  post-Newtonian timing formula.}
\newblock \emph{Ann.~Inst.~Henri Poincar{\'e} Phys.~Th{\'e}or.}, 44:\penalty0
  263--292, 1986.

\bibitem[{Blandford} and {Teukolsky}(1976)]{1976ApJ...205..580B}
R.~{Blandford} and S.~A. {Teukolsky}.
\newblock {Arrival-time analysis for a pulsar in a binary system.}
\newblock \emph{\apj}, 205:\penalty0 580--591, April 1976.
\newblock \doi{10.1086/154315}.

\bibitem[Shapiro(1964)]{PhysRevLett.13.789}
Irwin~I. Shapiro.
\newblock Fourth test of general relativity.
\newblock \emph{Phys. Rev. Lett.}, 13:\penalty0 789--791, Dec 1964.
\newblock \doi{10.1103/PhysRevLett.13.789}.
\newblock URL \url{https://link.aps.org/doi/10.1103/PhysRevLett.13.789}.

\bibitem[Abuter et~al.(2020{\natexlab{b}})]{2020S2}
R.~Abuter et~al.
\newblock Detection of the schwarzschild precession in the orbit of the star s2
  near the galactic centre massive black hole.
\newblock \emph{Astron. Astrophys.}, 636:\penalty0 L5, April
  2020{\natexlab{b}}.
\newblock ISSN 1432-0746.
\newblock \doi{10.1051/0004-6361/202037813}.
\newblock URL \url{http://dx.doi.org/10.1051/0004-6361/202037813}.

\end{thebibliography}
\bibliographystyle{unsrtnat}

%
\clearpage

\onecolumngrid

\setcounter{section}{0}
\renewcommand{\thesection}{\arabic{section}}
\setcounter{equation}{0}
\renewcommand{\theequation}{S\arabic{equation}}
\setcounter{figure}{0}
\renewcommand{\thefigure}{S\arabic{figure}}
\setcounter{table}{0}
\renewcommand{\thetable}{S\arabic{table}}
\begin{center}
\Large\bfseries Supplementary Materials
\end{center}

\section{Newtonian gravitational atom and spherical soliton}\label{sec2}
In the nonrelativistic (NR) limit, the slow mode of a massive spin-$s$ field can be described by a rank-$s$ tensor wavefunction $\psi_I$ in the Cartesian basis, here $I$ refers schematically to the set of tensor indices. E.g., for the scalar field, $\psi_I=\psi$; for the vector field, $I=i$; for the spin-2 filed, $I=ij$. Neglecting the possible non-gravitational self-interactions, the equation of motion of the wavefunction around a static point mass $M$ is given by the Shr\" {o}dinger-Poisson (SP) equation:
\begin{align}
i\partial_t\psi_I &=-\frac{1}{2m}\nabla^2\psi_I+m\left(\Phi-\frac{M}{r}\right) \psi_I,
\\
\nabla^2\Phi &=4\pi \rho,\quad \rho =m \sum_I|\psi_I|^2,
\end{align}
where $m$ is the boson mass, $\Phi$ is the Newtonian potential sourced by the wavefunction, corresponding to the weakly perturbed metric $ds^2=\left[-1-2\left(\Phi-\frac{M}{r}\right)\right]dt^2+\left[1-2\left(\Phi-\frac{M}{r}\right)\right]|d\mathbf{x}|^2$. In the following, we consider the bound state solution of this equation in the limiting cases of gravitational atom and spherically symmetric soliton. We denote the total mass of the bound state by $M_\text{c}=\int d^3r\,m\sum_I|\psi_I|^2\equiv \beta M$.

\subsection{Gravitational atom limit}
If $\beta \ll 1$, the self-gravity of the wavefunction can be neglected at the leading order. Setting $\Phi=0$, the SP equation reduces to the Shr\" {o}dinger equation of hydrogen atom:
\begin{equation}
	i\partial_t\psi_I=-\frac{1}{2m}\nabla^2\psi_I-\frac{\alpha}{r} \psi_I,
\end{equation}
where $\alpha\equiv m M$ is the gravitational fine-structure constant. For convenience, we introduce the Bohr radius $r_\text{c}\equiv M/\alpha^2$ and the dimensionless radial coordinate $x\equiv r/r_\text{c}=\alpha m r$. For each Cartesian component $I$, a general bound state can be written as
\begin{align}\label{general_bound}
\psi_I(t,r,\theta,\phi) &=\sum_{n\ge 1,\,
 0\le l\le n-1, \,|\mathtt{m}|\le l}c_{nl\mathtt{m}}\, \psi^{(nl\texttt{m})},
\\
\psi^{(nl\texttt{m})} &=r_\text{c}^{-3/2}R_{nl}(x)\,Y_{l\mathtt{m}}(\theta,\phi)\,e^{-i E^{(n)} t},
\end{align}
with $c_{nl\mathtt{m}}\in\mathbb{C}$, $E^{(n)} =-\frac{m\alpha^2}{2n^2}$, $Y_{l\mathtt{m}}(\theta,\phi)$ being the spherical harmonics, and
\begin{equation}
	R_{n l}(x)= \sqrt{\left(\frac{2}{n}\right)^3 \frac{(n-l-1) !}{2 n(n+l) !}}\left(\frac{2 x}{n}\right)^{l}e^{-\frac{x}{n}}\,L_{n-l-1}^{2 l+1}\left(\frac{2 x}{n}\right),
\end{equation}
where $\theta$ and $\phi$ are the angles written in the BH's coordinate system, $L^p_k(z)$ is the associated Laguerre polynomial.

For $\alpha \ll 1$, since $r_\text{c}\gg M$, this gravitational atom (GA) model provides an effective description for the superradiant cloud (quasi-bound states) around a spinning BH at radius $r\gg M$, if the $z$-axis is identified with the BH's spin direction. In the case of vector field, the NR limit of the quasibound states  turns out to be~\cite{Baryakhtar:2017ngi,B2,B3}
\begin{equation}
\psi_i(t,r,\theta,\phi)=\sum_{n\ge 1,\,0\le l\le n-1,\,|l-1|\le j\le l+1,\,|\texttt{m}|\le j} c_{nlj\texttt{m}}\,\psi_i^{(nlj\texttt{m})}.
\end{equation}
with
\begin{equation}
\psi_i^{(nlj\texttt{m})}=r_\text{c}^{-3/2}R_{nl}(x)\,Y^i_{lj\texttt{m}}(\theta,\phi)\,e^{-iE^{(n)}t},
\end{equation}
where the pure-orbital vector spherical harmonics\footnote{For the massive spin-2 field, the angular wavefunction is replaced by the pure-orbital tensor spherical harmonics~\cite{Brito_2020,RevModPhys.52.299} $Y_{lj\texttt{m}}^{ik}
	=\sum_{|m_s|\le 2}Y^{(m_s)}_{lj\texttt{m}}\,\xi_{ik}^{m_s} $, with $Y^{(m_s)}_{lj\texttt{m}}=\left\langle 2, m_s;l, \texttt{m}-m_s \mid j, \texttt{m}\right\rangle Y_{l,\texttt{m}-m_s}(\theta, \phi)$ and $\xi_{ik}^{m_s}=\sum_{|m_1|\le 1, |m_2|\le 1}\langle 1, m_1;1, m_2 \mid 2, m_s\rangle \,\xi_i^{m_1}\xi_k^{m_2}$. For $l=0$, $j=2$, and $Y_{02\texttt{m}}^{ik}=Y_{02\texttt{m}}^{(\texttt{m})}\,\xi^\texttt{m}_{ik}$. For $\alpha\ll1$, the superradiant ground state of spin-2 GA is $|nlj\texttt{m}\rangle=|1022\rangle$~\cite{Brito_2020}. The mass density distribution of the latter is the same as that of $|100\rangle$ state of scalar GA.} $Y^i_{lj\texttt{m}}=\sum_{|m_s|\le 1}Y^{(m_s)}_{lj\texttt{m}}\,\xi_i^{m_s}$ (with $\boldsymbol{\xi}^0=\mathbf{e}_z$ and $\boldsymbol{\xi}^{\pm 1}=\mp (\mathbf{e}_x\pm i \mathbf{e}_y)/\sqrt{2}$) has spherical components:
\begin{equation}
Y^{(m_s)}_{lj\texttt{m}}(\theta,\phi) = \langle 1,m_s;l,\texttt{m}-m_s|j,\texttt{m}\rangle \,Y_{l,\texttt{m}-m_s}(\theta,\phi),
\end{equation}
with $\langle j_1,m_1;j_2,m_2|j,m\rangle$ being the Clebsch-Gordan coefficients. $Y_{lj\texttt{m}}^i$ is the simultaneous eigenfunction of $\hat L_z$ and $|\hat{\mathbf{L}}|^2$, with the eigenvalues $\texttt{m}$ and $l(l+1)$, respectively; here $\hat{\mathbf{L}}\equiv \mathbf{r}\times(-i\nabla)$ is the orbital angular momentum operator. Note that  $\psi_i^{(nlj\texttt{m})}$ acquires a phase factor $(-1)^{l+1}$ under parity transformation, hence the $j=l\pm 1$ and $j=l$ modes are called electric and magnetic modes, respectively. For $l=0$, $j=1$, and  $Y^i_{01\texttt{m}}=Y^{(\texttt{m})}_{01\texttt{m}}\,\xi_i^{\texttt{m}}$.

The phenomenology of GA and GA-companion system has been extensively studied~\cite{Baryakhtar:2017ngi,Brito_2020,Ferreira:2017pth,B1,B2,B3,B4,Tomaselli:2023ysb,Tomaselli:2024bdd,Boskovic:2024fga,Tomaselli:2025jfo,Brito_2023,Duque:2023cac,dyson2025environmentaleffectsextrememass,Takahashi:2021yhy,Takahashi:2023flk,Zhang:2018kib,ZJ,Guo:2024iye,Liu:2024mzw,Peng:2025zca,Guo:2025ckp,WY_1,WY_2,WY_3,Fan:2023jjj,Kavic:2019cgk,DeLuca:2021ite,Arana:2024kaz,Su:2021dwz,Cao:2023fyv,Cao:2024wby,DellaMonica:2025zby,Lyu:2025lue,Guo:2025pea}. In this work, we consider the GA occupied by the fastest growing state $|nl\texttt{m}\rangle=|211\rangle$ for initial BH spin $\chi\sim 1$ (so-called superradiant ground state) in the case of scalar field, and $|nlj\texttt{m}\rangle=|1011\rangle$ in the case of vector field~\cite{B3}. The mass density distribution of the latter is the same as that of $|100\rangle$ state of scalar GA. We have
\begin{align}
\rho_{211} &= m^2\beta \alpha^4\frac{x^2e^{-x}\sin^2\theta}{64\pi}
,
\\
\rho_{100} &= m^2\beta \alpha^4\frac{e^{-2x}}{\pi}
.
\end{align}
Note that for $\beta\ll \alpha$,  $\rho/m^3 \sim \beta \alpha^4/m \ll \alpha^5/m \approx \left(\frac{M}{4.3\times 10^6M_\odot}\right)\left(\frac{\alpha}{0.1}\right)^4\times 2.1\,\text{ms}$.

Using the general solution to the Poisson equation $\nabla^2\Phi=4\pi \rho$:
\begin{align}
	\Phi(r,\theta,\phi)&=-\sum_{l\ge 0,\, |\mathtt{m}|\le l}\frac{4\pi}{2l+1}\left[\frac{q_{l\mathtt{m}}(r)}{r^{l+1}}+r^lp_{l\mathtt{m}}(r)\right]Y_{l\mathtt{m}}^*(\theta,\phi),\label{Poisson}
	\\
q_{l\mathtt{m}}(r)&=\int_0^rs^l\rho_{l\mathtt{m}}(s)s^2ds,
\\
p_{l\mathtt{m}}(r)&=\int_r^\infty\frac{\rho_{l\mathtt{m}}(s)}{s^{l+1}}s^2ds,
\\
\rho_{l\mathtt{m}}(r)&=\int_0^{2\pi}d\phi \int_0^\pi d\theta\,\sin\theta\int\rho(r,\theta,\phi)\, Y_{l\mathtt{m}}(\theta,\phi),\label{rho_sphere}
\end{align}
we obtain the Newtonian potential:
\begin{align}
	\frac{\Phi_{211}}{\beta \alpha^2}=&-\frac{1}{x}-\frac{3}{x^3}
	+\left(\frac{3}{x^3}+\frac{x^2}{16}+\frac{3}{x^2}+\frac{3 x}{8}+\frac{5}{2 x}+\frac{5}{4}\right) e^{-x}\nonumber
	\\
	&+\left[\frac{9}{x^3}+e^{-x} \left(-\frac{9}{x^3}-\frac{x^2}{16}-\frac{9}{x^2}-\frac{3 x}{8}-\frac{9}{2 x}-\frac{3}{2}\right)\right] \cos ^2\theta
	,
	\\
	\frac{\Phi_{100}}{\beta \alpha^2}=& -\frac{1}{x}+e^{-2 x} \left(\frac{1}{x}+1\right)\label{Phi_1011}
	.
\end{align}
The gravitational acceleration $\delta\mathbf{a}=-\nabla \Phi=-(\partial_r \Phi)\,\mathbf{e}_r-r^{-1}(\partial_\theta \Phi)\,\mathbf{e}_\theta$ is given by
\begin{align}
    \frac{\delta \mathbf{a}_{211}}{\beta \alpha^4/M} =& \Big\{-\frac{1}{x^2}+\frac{e^{-x}}{16x^4}\Big[144-144e^x+144x+88x^2+40x^3+14x^4+4x^5+x^6\nonumber\\
    &+\left(-432+432e^x-432x-216x^2-72x^3-18x^4-4x^5-x^6\right)\cos^2\theta\Big]\Big\}\mathbf{e}_r\nonumber\\
    &+\frac{e^{-x}}{8x^4}\Big[\left(-144+144e^x-144x-72x^2-24x^3-6x^4-x^5\right)\cos \theta \sin \theta\Big]\mathbf{e}_{\theta},
    \\
       \frac{\delta \mathbf{a}_{100}}{\beta \alpha^4/M} =& \frac{e^{-2 x}}{x^2}\left[1-e^{2x}+2x(1+x)\right]\mathbf{e}_r.
\end{align}

Here we neglect the relativistic corrections, e.g., from the gravitomagnetic field of the cloud due to its angular momentum~\cite{Cao:2024wby}. At leading order in $\alpha$, the stationary gravitomagnetic potential of the $|211\rangle$ and $|1011\rangle$ states are given by
\begin{align}
\boldsymbol{\Xi} =-\frac{2 J_\text{c} \sin\theta}{r^2}\mathbf{e}_\phi
,
\quad
J_\text{c}(\mathbf{r}) =\frac{M_\text{c}}{\mu}\mathcal{J}(x),
\end{align}
with
\begin{align}
\mathcal{J}_{211}(x) & =1-\frac{1}{8}e^{-x}(x+2)(x^2+2x+4)
,
\\
\mathcal{J}_{1011}(x) & =1 - e^{-2x}(1+2x+2x^2)
.
\end{align}
The gravitomagnetic acceleration $\mathbf{v}\times (\nabla\times\boldsymbol{\Xi})$ is suppressed by $\mathcal{O}(\alpha|\mathbf{v}|)$ relative to $-\nabla\Phi$. Meanwhile, dissipative effects are suppressed by the mass ratio $M_*/M$, where $M_*$ is the mass of the companion. Despite their possible relevance for the secular orbital evolution, they are much weaker than the conservative effects if the cloud is only weakly perturbed~\cite{Cao:2024wby}, and consequently are also neglected.\footnote{Although the dissipative effects are negligible for short-term evolution, they can potentially modify the distribution of orbital parameters in the nuclear star cluster, as recently discussed in \cite{tomaselli2025probingdenseenvironmentssgr}.}

\subsection{Spherically symmetric ground state}
Even if $\alpha \ll 1$, the GA approximation breaks down if $\beta$ becomes sufficiently large. In this section, we consider the exact spherically symmetric solution to the SP equation, focusing particularly on the ground state. Note that the solution taking into account the central mass has also been studied in \cite{Chavanis:2019bnu,Bar:2019pnz,Davies_2020}.

Without loss of generality, we consider the case of scalar field.\footnote{At the level of the SP equation, a spherically symmetric mass distribution of higher-spin fields can still possess spin angular momentum. In these cases, we again neglect the resulting gravitomagnetic fields.} For the spherically symmetric ansatz:
\begin{equation}
\psi(t,r)=f(r)\,e^{-iEt},
\end{equation}
with $f\in\mathbb{R}$, the SP equation is reduced to
\begin{equation}\label{eq1}
Ef=-\frac{1}{2m}\frac{\partial_r\left(r^2 \partial_r f\right)}{r^2} + m \left(\Phi-\frac{M}{r}\right) f
,\quad
\frac{\partial_r\left(r^2 \partial_r\Phi\right)}{r^2}=4\pi m f^2.
\end{equation}
Introducing the nondimensional radius $y\equiv m r$ and
\begin{equation}
V=2\left(\Phi-\frac{E}{m}\right),
\quad
F=\sqrt{\frac{8\pi}{m}}\,f,
\end{equation}
Eq.~\eqref{eq1} is further simplified to
\begin{equation}\label{eq2}
y \partial_y^2F+2\partial_yF=(yV-2\alpha )F
,
\quad
y\partial_y^2V+2\partial_y V=yF^2.
\end{equation}
A bound state ($E<0$) solution $\{V_\kappa(y,\alpha),F_\kappa(y,\alpha)\}$ can be specified by the boundary condition $F(0,\alpha)= \kappa^2$ with $\partial_yV(0,\alpha)=\partial_yF(0,\alpha)=F(\infty,\alpha)=0$. For a given number of nodes in the wavefunction, the value of $V_\kappa(0,\alpha)$ can be determined numerically by the shooting method.

\begin{figure}[hbt!]
\includegraphics[width=0.45\textwidth]{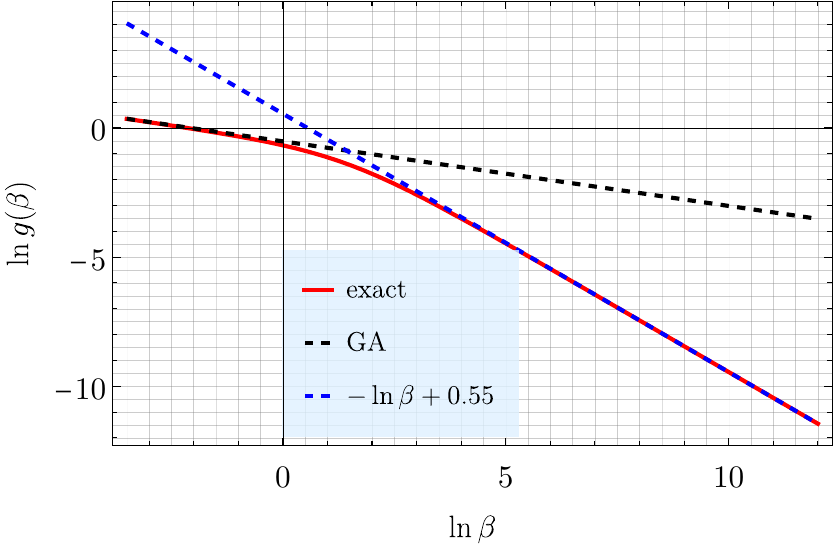}
\caption{$g(\beta)$.}
\label{fig:g_beta}
\end{figure}

Eq.~\eqref{eq2} has a scaling symmetry:
\begin{align}
F_\kappa(y,\alpha) &=\kappa^2 F_1(\kappa y,\alpha/\kappa),
\\
V_\kappa(y,\alpha) &=\kappa^2 V_1(\kappa y,\alpha/\kappa),
\end{align}
it follows that
\begin{align}
\beta_\kappa(\alpha) &=\beta_1(\alpha/\kappa)= B(\alpha/\kappa),
\\
\Phi_\kappa(y,\alpha) &=\kappa^2 \Phi_1(\kappa y,\alpha/\kappa),
\\
E_\kappa(\alpha,\gamma) &=\kappa^2 E_1(\alpha/\kappa)=\left(-\frac{m\alpha^2}{2}\right)C(\beta).\label{E_kappa}
\end{align}
where we define the functions:
\begin{align}
B(\alpha) &\equiv \frac{1}{2\alpha}\int_0^\infty dy\,y^2\,[F_1(y,\alpha)]^2,
\\
C(\beta) &\equiv V_1(\infty,g(\beta))\,[g(\beta)]^{-2}.
\end{align}
Therefore we only need to solve the bound state for $\kappa=1$. The $\alpha$-$\beta$ relation can be obtained from $\beta=B(\alpha/\kappa)$ as $g(\beta)=\alpha/\kappa$, with $g(x)$ being the inverse function of $B(x)$ (see Fig.~\ref{fig:g_beta}). Utilizing the scaling relations above, the mass density profile is explicitly given by
\begin{equation}
\begin{aligned}
\rho & = m|\psi|^2=
\frac{m^2}{8\pi}\left[\frac{\alpha}{g(\beta)}\right]^4 \left[F_1\left(\frac{\alpha m r}{g(\beta)},g(\beta)\right)\right]^2
.
\end{aligned}
\end{equation}
Note that $\alpha m r=x=r/r_\text{c}$, the normalized density profile thus depends only on $\beta$ and $x$.

\begin{figure}[hbt!]
\includegraphics[width=0.45\textwidth]{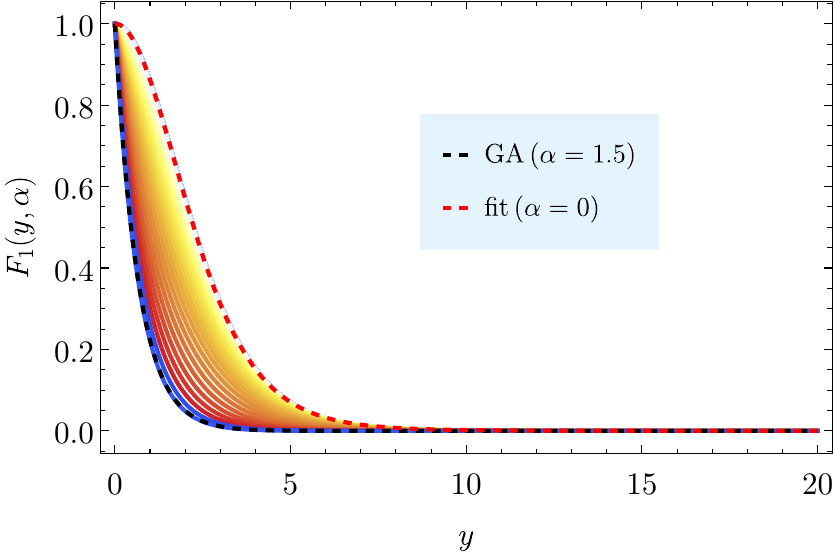}
\;
\includegraphics[width=0.45\textwidth]{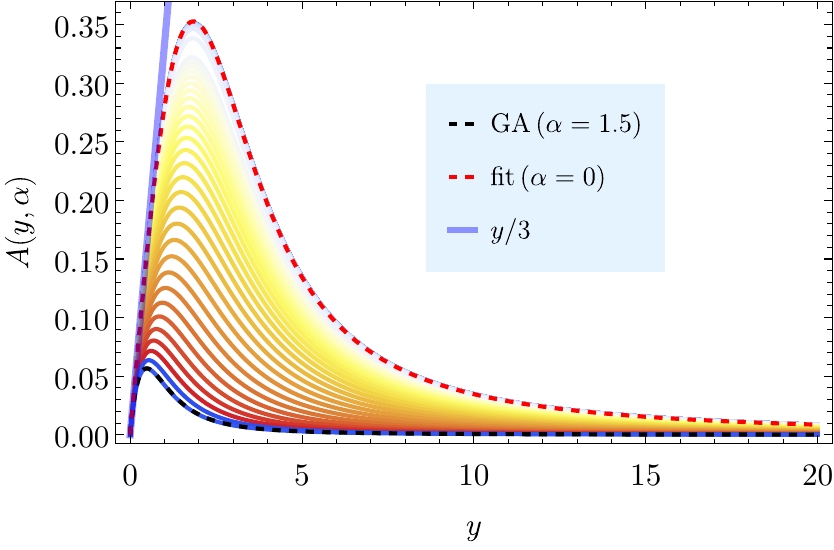}
\caption{The function $F_1(y,\alpha)$ characterizing the radial profile of the ground state, and the function $A(y,\alpha)$ characterizing the gravitational acceleration.}
\label{fig:A_and_F1}
\end{figure}

Correspondingly, the gravitational acceleration is given by $\delta \mathbf{a}=\delta a\,\mathbf{e}_r$, with
\begin{equation}
\begin{aligned}
\delta a(r)=-\frac{m}{2}\left[\frac{\alpha}{g(\beta)}\right]^3 A\left(\frac{\alpha m r}{g(\beta)},g(\beta)\right)
,
\end{aligned}
\end{equation}
and $A(y,\alpha)\equiv \partial_y V_1(y,\alpha)$ (see Fig.~\ref{fig:A_and_F1}). From this we also obtain the enclosed mass profile as $M_\text{c}(r)=-r^2\delta a(r)$.

\begin{figure}[hbt!]
\includegraphics[width=0.45\textwidth]{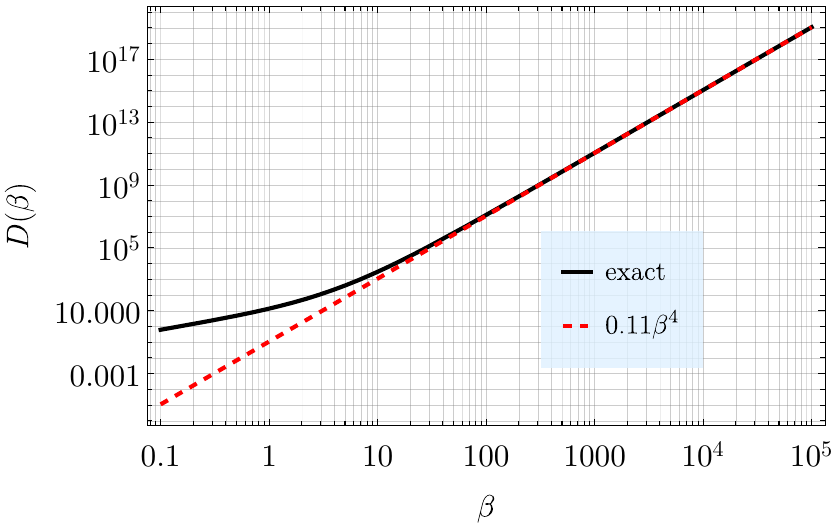}
\caption{The function $D(\beta)$ characterizing the central mass density of the ground state.}
\label{fig:D_beta}
\end{figure}

The ground state is the state with minimum value of $E$ and is characterized by the absence of nodes in the wavefunction. Some limiting behaviours of the ground state are as follows:
\begin{itemize}

\item  For $y\to 0$, $V_1(y,\alpha)\to y^2/6$, thus $A(y,\alpha)\to y/3$, independent of $\alpha$. The density profile is given by $M_\text{c}(r)/M \to (4\pi/3)\,\rho(0)\, r^3/M=m^6\,D(\beta)\,(Mr)^3/6$, with $D(\beta)\equiv \left[g(\beta)\right]^{-4}\left[F_1\left(0,g(\beta)\right)\right]^2$ (see Fig.~\ref{fig:D_beta}). For $\beta \gtrsim 100$, $D(\beta)\approx 0.11\beta^4$.

\item For $y\to \infty$, $\delta a\to -\beta M/r^2$, and $A(y,\alpha)\to 2B(\alpha)\,\alpha/y^2$.

\item For $\alpha=0$ (i.e., without the central point mass), a good fit is given by (compatible with \cite{Schive:2014dra})
\begin{equation}
F_1(y,0)=\frac{1}{(1+0.03762\,y^2)^4}.
\end{equation}

\item For $\beta \to 0$, the ground state tends smoothly to the $|100\rangle $ state of scalar GA, with
\begin{equation}
	F_1(y,\alpha)=e^{-\alpha y},\quad 
	V_1(0,\alpha)=\alpha^2,\quad
	B(\alpha)=\frac{1}{8 \alpha ^4},
\quad
	g(\beta)=(8\beta)^{-1/4},
\end{equation}
and
\begin{equation}
	A(y,\alpha)=\frac{1-e^{-2  \alpha y}[1+2 \alpha y (1+\alpha y)]}{4 \alpha^3y^2}.
\end{equation}

\end{itemize}

Note that for $\beta \gtrsim 100$, $\rho/m^3<\rho(0)/m^3=\frac{M}{8\pi}\alpha^3D(\beta) \approx \left(\frac{M}{4.3\times 10^6M_\odot}\right)\left(\frac{\alpha}{0.1}\right)^3\beta^4\times 0.1\,\text{ms}$.

\begin{figure}[hbt!]
\includegraphics[width=0.45\textwidth]{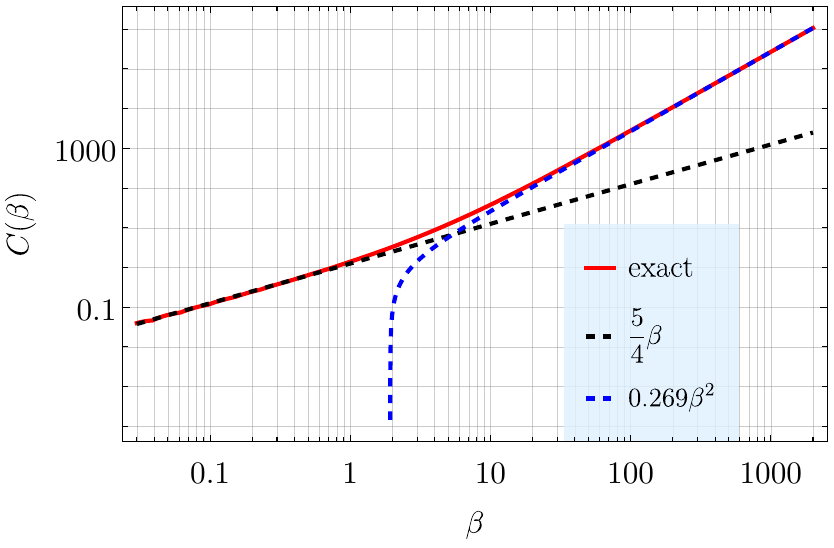}
\caption{The function $C(\beta)$ characterizing the energy level of the ground state.}
\label{fig:C_beta}
\end{figure}

For a sufficiently small value of $\beta$, the ground state can be approximated by a scalar $|100\rangle $ state with self-gravity correction, we take this chance to check its range of validity. Using Eqs.~\eqref{Phi_1011} and \eqref{E_kappa}, the correction to the energy level is
\begin{equation}
\langle 100|m \Phi_{100}| 100\rangle=\int d^3r\,\left|\psi^{100}\right|^2 m \Phi_{100} = -\frac{5}{8}m\beta \alpha^2,
\end{equation}
thus
\begin{equation}
C(\beta)\approx\frac{E^{(1)}+\langle 100|m \Phi_{100}| 100\rangle}{-m\alpha^2/2}=1+\frac{5}{4}\beta.
\end{equation}
As shown in Fig.~\ref{fig:C_beta}, this approximation remains good even for $\beta \sim 1$, while $C(\beta)\propto \beta^2$ in the large-$\beta$ limit.

\begin{figure}[hbt!]
\includegraphics[width=0.45\textwidth]{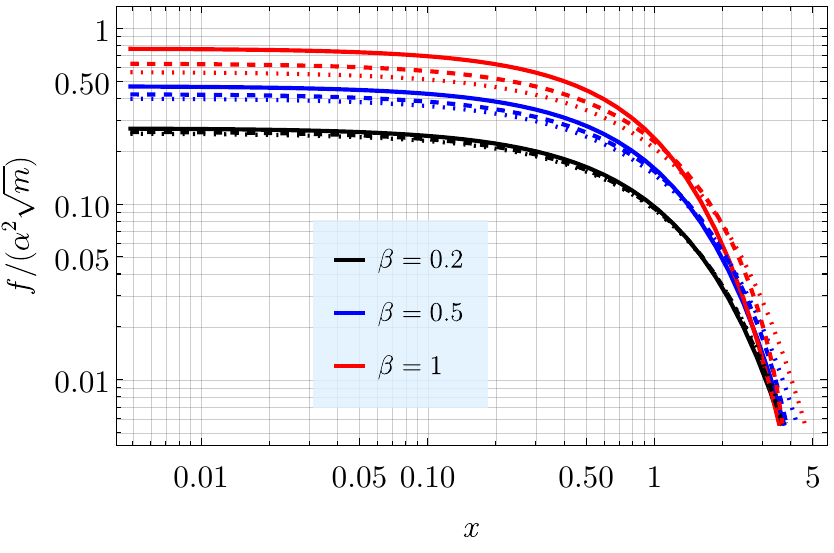}
\caption{Comparison between the exact radial wavefunction of self-gravitating spherically symmetric ground state (solid line), the GA limit (dotted line) and the result with self-gravity correction (dashed line), for different values of $\beta$.}
\label{fig:wave_function_correction}
\end{figure}

The leading-order correction to the wavefunction of $|100\rangle$ state comes from its mixings with the $|n\ge 2,00\rangle$ states, given by
\begin{equation}
\Delta\psi^{(100)}=\sum_{n\ge 2}\frac{\langle n00|m\Phi_{100}|100\rangle}{E^{(1)}-E^{(n)}}\psi^{(n00)}.
\end{equation}
Compared with the exact radial wavefunction
\begin{equation}
\frac{f(r)}{\alpha^2\sqrt{m}}=\frac{1}{\sqrt{8\pi}\,[g(\beta)]^2} F_1\left(\frac{x}{g(\beta)},g(\beta)\right),
\end{equation}
the corrected radial wavefunction takes the form
\begin{equation}
\frac{f(r)}{\alpha^2\sqrt{m}}=\sqrt{\beta }\left[\frac{e^{-x}+\beta \mathcal{F}(x)}{\sqrt{\pi }}\right].
\end{equation}
The function $\mathcal{F}(x)$ after including the mixing with $n=2,3,4$ $s$-states is
\begin{equation}
\mathcal{F}(x)\approx -\frac{8192 e^{-\frac{x}{2}} (x-2)}{194481}+\frac{141 e^{-\frac{x}{3}} [2 (x-9) x+27]}{200000}-\frac{2891776 e^{-\frac{x}{4}} [(x-12)^2 x-192]}{75418890625},
\end{equation}
and the contribution from higher-$n$ states is small. As shown in Fig.~\ref{fig:wave_function_correction}, this correction is small for $\beta<1$, while for $\beta\gtrsim 1$ the approximation becomes poor.

\section{Quasibound states of massive scalar and vector fields in Kerr spacetime} \label{sec:appendixB}
In this section, we compare the nonrelativistic GA approximation with the relativistic quasibound state (QBS) solutions for the scalar $|211\rangle$ and vector $|1011\rangle$ states.

The relativistic field profiles associated with the NR wavefunction is given by
\begin{equation}
\phi=\frac{1}{\sqrt{2m}}\left(\psi\, e^{-im t}+\text{c.c.}\right),
\end{equation}
for the real\footnote{In the test-field limit, the situation is similar for a complex field, which can be represented by $\phi=\frac{1}{\sqrt{2m}}\left(\psi_+\, e^{-im t}+\psi_-\, e^{-im t}\right)$.} scalar field $\phi$, and
\begin{equation}
A_i=\frac{1}{\sqrt{2m}}\left(\psi_i\, e^{-im t}+\text{c.c.}\right),
\end{equation}
for the real vector field $A_i$, such that the energy density $-T^0_0$ reduces to the mass density $\rho=m\sum_I|\psi_I|^2$ in the NR limit. For the GA occupied by a single state, we thus have
\begin{align}
\phi^\text{(GA)}_{nl\texttt{m}}(t,r,\theta,\phi) &=\sqrt{\frac{M_\text{c}}{2m^2}} \left[\psi^{(nl\texttt{m})}\, e^{-im t}+\text{c.c.}\right],
\\
\mathbf{A}^{\text{(GA)}}_{nlj\texttt{m}}(t,r,\theta,\phi) &= \sqrt{\frac{M_\text{c}}{2m^2}} \left[\boldsymbol{\psi}^{(nlj\texttt{m})}\, e^{-im t}+\text{c.c.}\right].
\end{align}
Using
\begin{equation}
R_{21}(x)=\frac{e^{-x/2} x}{2 \sqrt{6}},
\quad
Y_{11}(\theta,\phi)=-\sqrt{\frac{3}{8 \pi }} e^{i \phi } \sin \theta
,
\end{equation}
and
\begin{equation}
R_{10}(x)=2 e^{-x},
\quad
Y_{00}(\theta,\phi)=\frac{1}{\sqrt{4\pi }}
,
\end{equation}
we obtain for the scalar $|211\rangle$ state,
\begin{equation}
\phi_{211}^\text{(GA)}(t,r,\theta,\phi)=-\frac{\alpha ^2 \sqrt{\beta } }{4 \sqrt{2 \pi }}e^{-x/2} x \sin \theta\, \cos (\omega_{211} t- \phi),
\end{equation}
and for the vector $|1011\rangle$ state (in the spherical coordinates),
\begin{equation}
\left(\begin{matrix}
A_{1011}^{\text{(GA)}t}
\\
A_{1011}^{\text{(GA)}r}
\\
A_{1011}^{\text{(GA)}\theta}
\\
A_{1011}^{\text{(GA)}\phi}
\end{matrix}\right)
=
\left(\begin{matrix}
A_{1011}^{\text{(GA)}t}
\\
\mathbf{A}_{1011}^{\text{(GA)}}\cdot\mathbf{e}_r
\\
\mathbf{A}_{1011}^{\text{(GA)}}\cdot\mathbf{e}_\theta/r
\\
\mathbf{A}_{1011}^{\text{(GA)}}\cdot\mathbf{e}_\phi/(r\sin\theta)
\end{matrix}\right)
=-\frac{\alpha ^2 \sqrt{\beta }}{\sqrt{\pi }} e^{-x}\left(\begin{matrix}
 -\alpha \sin \theta\, \sin \left(\omega_{1011} t-\phi \right)
 \\
\sin \theta\, \cos \left(\omega_{1011} t-\phi \right)
 \\
\cos \theta\, \cos \left(\omega_{1011} t-\phi \right)
 \\
\sin \left(\omega_{1011} t-\phi \right)
\end{matrix}\right),
\end{equation}
with $ \omega_{211}=m+E^{(2)}$, $ \omega_{211}=m+E^{(1)}$ and $x=r/r_\text{c}$.

The superradiant cloud produced by a spinning black hole (in the test field limit) lives in the Kerr spacetime, and the GA approximation above unavoidably breaks down in the inner region sufficiently close to the event horizon. To check this deviation, we compute the QBSs of free scalar and vector fields corresponding to the GA states $|211\rangle$ and $|1011\rangle$ in the NR limit.

In the Boyer-Lindquist coordinates $(t,r,\theta,\phi)$, the metric of Kerr spacetime with mass parameter $M$ and spin parameter $a=M\chi$ is given by
\begin{equation}
	g_{a b}=\left(\begin{smallmatrix}
		-1+\frac{2 M r}{r^2+a^2 \cos ^2 \theta} & 0 & 0 & -\frac{2 M r a \sin ^2 \theta}{r^2+a^2 \cos ^2 \theta} \\
		0 & \frac{r^2+a^2 \cos ^2 \theta}{r^2-2 M r+a^2} & 0 & 0 \\
		0 & 0 & r^2+a^2 \cos ^2 \theta & 0 \\
		-\frac{2 M r a \sin ^2 \theta}{r^2+a^2 \cos ^2 \theta} & 0 & 0 & \sin ^2 \theta\left(r^2+a^2+\frac{2 M r a^2 \sin ^2 \theta}{r^2+a^2 \cos ^2\theta}\right)
	\end{smallmatrix}\right).
\end{equation}

The seperable ansatz of a free real scalar field is
\begin{equation}
\phi^\text{(rel)}_{nl\texttt{m}}(t,r,\theta,\phi)\propto e^{-i\omega t+i\texttt{m}\phi}S(\theta)\,R(r)+\text{c.c.},
\end{equation}
where $S=S_{l\texttt{m}}(\gamma,\theta)$ is the spheroidal harmonics with the oblateness parameter $\gamma=ia\sqrt{\omega^2-\mu^2}$. Quasibound states correspond to solutions that are purely ingoing at the horizon $r_+/M=1+\sqrt{1-\chi^2}$ and exponentially decaying at infinity, this results in a complex discrete spectrum: $\omega=\omega_R+i\omega_I$, with $\omega_R<m$ and $\omega_I\propto \texttt{m}\Omega_H-\omega_R$, $\Omega_H=\frac{1}{2M}\frac{\chi}{1+\sqrt{1-\chi^2}}$ being the angular velocity of the horizon. The superradiant growth ($\omega_I>0$) happens when $\texttt{m}\Omega_H>\omega_R$. Note that $n=l+1+\hat n_S$, with the overtone number $\hat n_S\ge 0$. We compute the $|211\rangle$ QBS numerically following the approach of \cite{Dolan_2007}. See also \cite{B2,Bao_2022} for the analytical approximations.

The seperable ansatz of a free real vector field (for all electric modes and a subset of magnetic modes \cite{B2}) is \cite{Frolov:2018ezx}
\begin{equation}
A^{\text{(rel)}\,a}_{nlj\texttt{m}}(t,r,\theta,\phi)=B^{ab}\,\partial_bZ,
\quad
Z\propto e^{-i\omega t+i\texttt{m}\phi}S(\theta)\,R(r)+\text{c.c.},
\end{equation}
where $B^{ab}(\nu)$ is related to an eigenvalue $\nu$ by $B^{ab}(g_{bc}+i\nu h_{bc})=\delta^a_c$, and $h_{ab}$ is the principal Killing–Yano tensor of Kerr spaceime satisfying $\nabla_ah_{bc}=g_{ab}\xi_c-g_{ac}\xi_b$, with $\xi^c$ being the timelike Killing vector. In the Boyer-Lindquist coordinates, $\xi^c=\partial_t x^c$,
\begin{equation}
h_{ab}=\left(\begin{smallmatrix}0&r&a^2\cos\theta\sin\theta&0\\-r&0&0&ar\sin\theta^2\\-a^2\cos \theta\sin\theta&0&0&a\cos\theta \sin\theta\,(a^2+r^2)\\0&-ar\sin\theta^2&-a\cos\theta\sin\theta\,(a^2+r^2)&0\end{smallmatrix}\right).
\end{equation}
The meaning of $\{n,l,j,\texttt{m}\}$ comes from the $\alpha\to 0$ limit, with the correspondence $j=l-S$ and $|S|\le 1$. Note that in the convention of \cite{PhysRevD.98.104006}, $n=|\texttt{m}|+1+S+\hat n_V$, with the overtone number $\hat n_V\ge 0$. We compute the $|1011\rangle$ QBS using the numerical solver provided by \cite{Fell_2023}.

\begin{figure}[hbt!]
\includegraphics[width=0.45\textwidth]{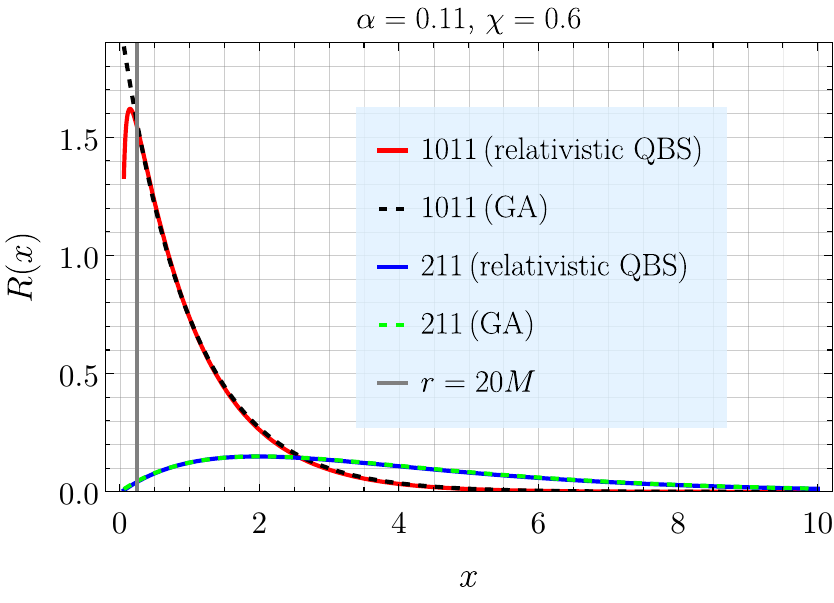}
\caption{Comparison between the radial profiles of GA and relativistic QBS for the scalar $|211\rangle$ and vector $|1011\rangle$ states.}
\label{fig:compare_relativistic}
\end{figure}

Fig.~\ref{fig:compare_relativistic} displays the results for $\alpha=0.11$ and $\chi=0.6$, where for concreteness we compare the radial functions $R_{10}(x)$ and $R_{21}(x)$ with $\left[\frac{R_{10}(1)}{A_{1011}^{\text{(rel)}\,r}(0,1,\pi/2,0)}\right]A_{1011}^{\text{(rel)}\,r}(0,x,\pi/2,0)$ and $\left[\frac{R_{21}(1)}{R_{211}^{\text{(rel)}}(1)}\right]R_{211}^{\text{(rel)}}(x)$, respectively. The agreement between GA and QBS profiles appears to be good at $r> 20M$ (this holds also for the angular profiles), and is better for smaller values of $\alpha$. This justifies our nonrelativistic modeling of the bosonic field and its metric perturbation at large radius, when the self-gravity of the field is negligible.\footnote{In this regime, the (non-radiative) linear metric perturbation sourced by the QBS can, in principle, be computed within black-hole perturbation theory, which would provide a more accurate description of its gravitational effects, particularly on the the conservative dynamics of a small companion.}

\section{Orbital dynamics}\label{orbit}


\begin{figure*} 
\includegraphics[width=0.4\textwidth]{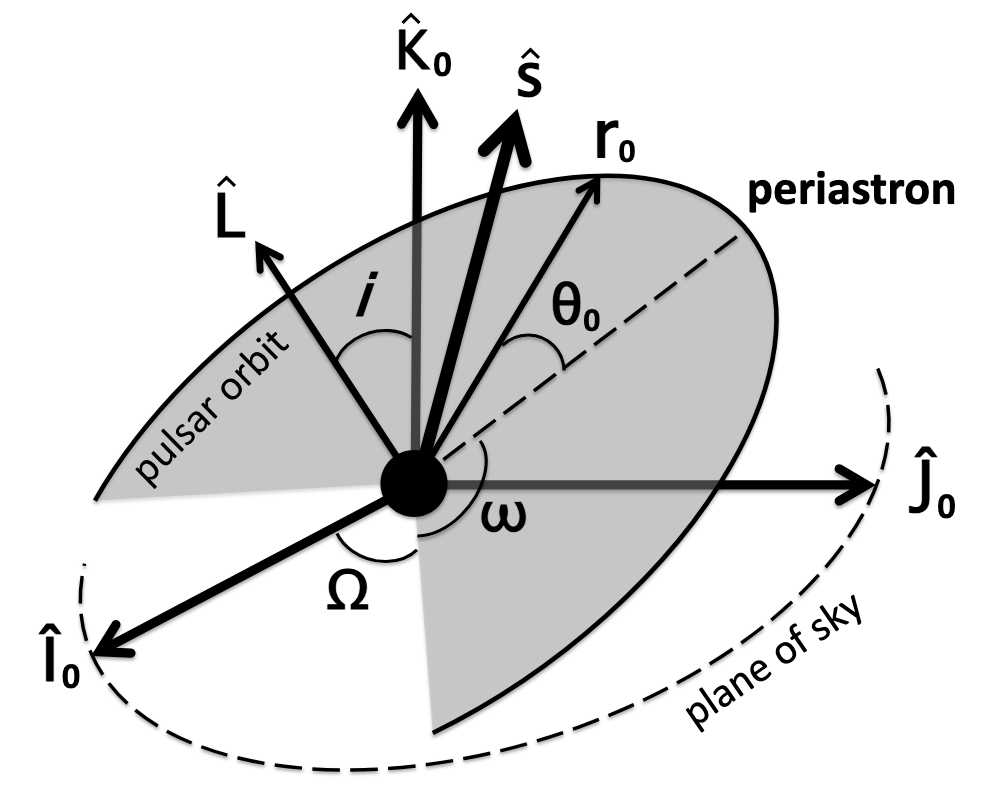}
\caption{Schematic of a pulsar orbit around Sgr A$^{*}$.}
\label{fig:frame}
\end{figure*}

In this work, we use PN expansion to solve for the two-body orbit, the relative acceleration can be expressed as
\begin{align}\label{acc}
    \ddot{\mathbf{r}}\equiv\frac{\mathrm{d}^2\mathbf{r}}{\mathrm{d}t^2}=\ddot{\mathbf{r}}_\text{N}+\ddot{\mathbf{r}}_\text{1PN}+\ddot{\mathbf{r}}_\text{SO}+\ddot{\mathbf{r}}_\text{Q}+\ddot{\mathbf{r}}_\text{2PN}+\ddot{\mathbf{r}}_\text{DM},
\end{align}
where $\mathbf{r}$ is a vector which refers to the relative coordinate position in the harmonic gauge, $t$ is the coordinate time, and the terms to the right of the equal sign represent Newtonian acceleration, 1PN acceleration, spin-orbit coupling acceleration, quadrupolar acceleration, 2PN acceleration and acceleration induced by ultralight field's gravitational potential, respectively. Here we ignore the higher PN terms.

Because the mass of the pulsar is far smaller than the mass of BH, the mass ratio $m_{\text{PSR}}/M < 10^{-6}$, we can ignore the mass of the pulsar and treat the pulsar as a test particle moving in the BH's spacetime. In this case, the terms in Eq.~\eqref{acc} are given by \cite{Hu:2023ubk}
\begin{align}
    \ddot{\mathbf{r}}_\text{N}&=-\frac{M}{r^2}\mathbf{e}_r,\\
    \ddot{\mathbf{r}}_\text{1PN}&=-\frac{M}{r^2}\left[\left(-\frac{4M}{r}+v^2\right)\mathbf{e}_r-4\dot{r}\mathbf{v}\right], \label{1PN}\\
    \ddot{\mathbf{r}}_\text{SO}&=\chi\frac{6M^2}{r^3}\left[\hat{\mathbf{s}}\cdot(\mathbf{e}_r\times \mathbf{v})\mathbf{e}_r+\dot{r}(\mathbf{e}_r\times \hat{\mathbf{s}})-\frac{2}{3}(\mathbf{v}\times \hat{\mathbf{s}})\right],\\
    \ddot{\mathbf{r}}_\text{Q}&=-q\frac{3M^3}{2r^4}\left\{\left[5(\mathbf{e}_r\cdot \hat{\mathbf{s}})^2-1\right]\mathbf{e}_r-2(\mathbf{e}_r\cdot \hat{\mathbf{s}})\hat{\mathbf{s}}\right\},\\
    \ddot{\mathbf{r}}_\text{DM}&= \delta \mathbf{a},
\end{align}
where $r\equiv|\mathbf{r}|$, $\dot{r}\equiv \mathrm{d}r/\mathrm{d}t$, $\mathbf{v}\equiv \mathrm{d}\mathbf{r}/\mathrm{d}t$ and $v\equiv|\mathbf{v}|$. 

The Keplerian orbit of the pulsar and the corresponding coordinate system and notations are shown in Fig.~\ref{fig:frame}. The $\mathbf{K}_0$ represent the direction from Earth to Sgr~A*, and $(\mathbf{I}_0,\mathbf{J}_0)$ forms the sky plane. The orbital parameters of pulsars include orbital period $P_b$, eccentricity $e$, inclination $i$, longitude of the periastron $\omega$, longitude of ascending node $\Omega$ and initial true anomaly $\theta_0$. $\eta$ and $\lambda$ represent the spin orientation of the BH: $\hat{\mathbf{s}}=(\sin \lambda \cos \eta, \sin \lambda \sin \eta, \cos \lambda)$. In this simulation, the longitude of the ascending node ($\Omega$) is fixed at $\Omega = 0$, and the other parameters are listed in Table~\ref{tab:benchmark}.

\begin{table}[h!]
\centering
\begin{tabularx}{\textwidth}{@{\extracolsep{\fill}} cccccccccc }
\hline
$M/M_\odot$ & $\chi$  & $q$  & $\lambda$ & $\eta$ & $P_b$ & $e$  & $i$  & $\omega$ & $\theta_0$\\
\hline
$4.3\times 10^6$ & $0.6$  & $-0.36$  & $\frac{1}{3}\pi$ & $\frac{5}{9}\pi$ & $0.5\,\text{yr}$ & $0.8$  & $\frac{1}{5}\pi$  & $\frac{5}{7}\pi$ & $\frac{1}{3}\pi$\\
\hline
\end{tabularx}
\caption{Parameters used in the simulation.}
\label{tab:benchmark}
\end{table}


The observation time in our simulation is 5 years, that contains $\sim 10$ orbits. Apart from that, we also considered cases with orbital period of $P_b\sim 5$ years and eccentricity of $e \sim 0.3$ to compare the effects of ULDM on different pulsar orbits. 

It is worth mentioning that, in order to express the acceleration caused by ULDM in the scene we defined above, we need to calculate $\theta$ and  $\mathbf{e}_{\theta}$ with $\cos \theta = \hat{\mathbf{s}}\cdot \mathbf{e}_r$ and 
\begin{equation}
    \mathbf{e}_{\theta}=\frac{(\hat{\mathbf{s}}\times\mathbf{e}_r)\times\mathbf{e}_r}{|\hat{\mathbf{s}}\times\mathbf{e}_r|}.
\end{equation}

\section{Pulsar timing}\label{pt}
\subsection{Timing model}
To explore the possibility of constraining the bosonic field near Sgr A$^*$ using pulsar timing, the first step is to develop a timing model that properly accounts for the extra attraction from the bosonic field stucture. In pulsar timing, the model connects the pulsar’s intrinsic rotation to the observed times of arrival (TOAs) at radio telescopes by incorporating various physical effects that influence the signal during its propagation from the pulsar to the Earth.

In pulsar's frame, the proper rotation numbers N of the pulsar can be expressed by
\begin{align}
N(T)=N_0+\nu\,T+\frac{1}{2}\dot{\nu}\,T^2,
\end{align}
where $\nu \equiv 1/P$ is the spin frequency, $T$ is the proper time of the pulsar, and $\dot{\nu}$ is the time derivative of $\nu$. The relation between the TOAs in the observer frame and the proper time is \cite{AIHPA_1986__44_3_263_0}
\begin{align}
    t^{\text{TOA}}=T+\Delta_E+\Delta_R+\Delta_S,
\end{align}
where $\Delta_E, \Delta_R, \Delta_S$ refer to Einstein delay, Romer delay and Shapiro delay. First, the Einstein delay is defined by the difference between the coordinate time $t$ and the proper time $T$ \cite{AIHPA_1986__44_3_263_0,1976ApJ...205..580B}:
\begin{align}\label{einlay}
    \Delta_E \equiv t - T.
\end{align}
At the lowest order, the proper time $T$ is connected to $t$ through

\begin{align}
    \frac{\mathrm{d}T}{\mathrm{d}t}=1-\frac{M}{r}-\frac{v^2}{2}.
\end{align}
We can see from above that the Einstein delay includes both gravitational redshift and special-relativistic time-dilation effects. The next one is Romer delay $\Delta_R$, which is a geometric effect caused by the orbital motion of the pulsar. Its form is
\begin{align}
    \Delta_R \equiv \hat{\mathbf{K}}_0 \cdot \mathbf{r}.
\end{align}
The last term in our timing model is 1PN Shapiro delay. Its existence is due to the propagation of light in curved spacetime \cite{1976ApJ...205..580B,PhysRevLett.13.789}:
\begin{align}\label{shalay}
    \Delta_S=-2M \ln \left(r-\mathbf{r}\cdot \hat{\mathbf{K}}_0\right).
\end{align}
In our timing model, we assume that the form of the Einstein delay and Shapiro delay in Eqs.~\eqref{einlay} and \eqref{shalay} do not include the ULDM's effect, since it is negligible compared to the Romer delay. 

\subsection{Timing residual induced by the ULDM}\label{tr}
\begin{figure*} 
\includegraphics[width=0.55\textwidth]{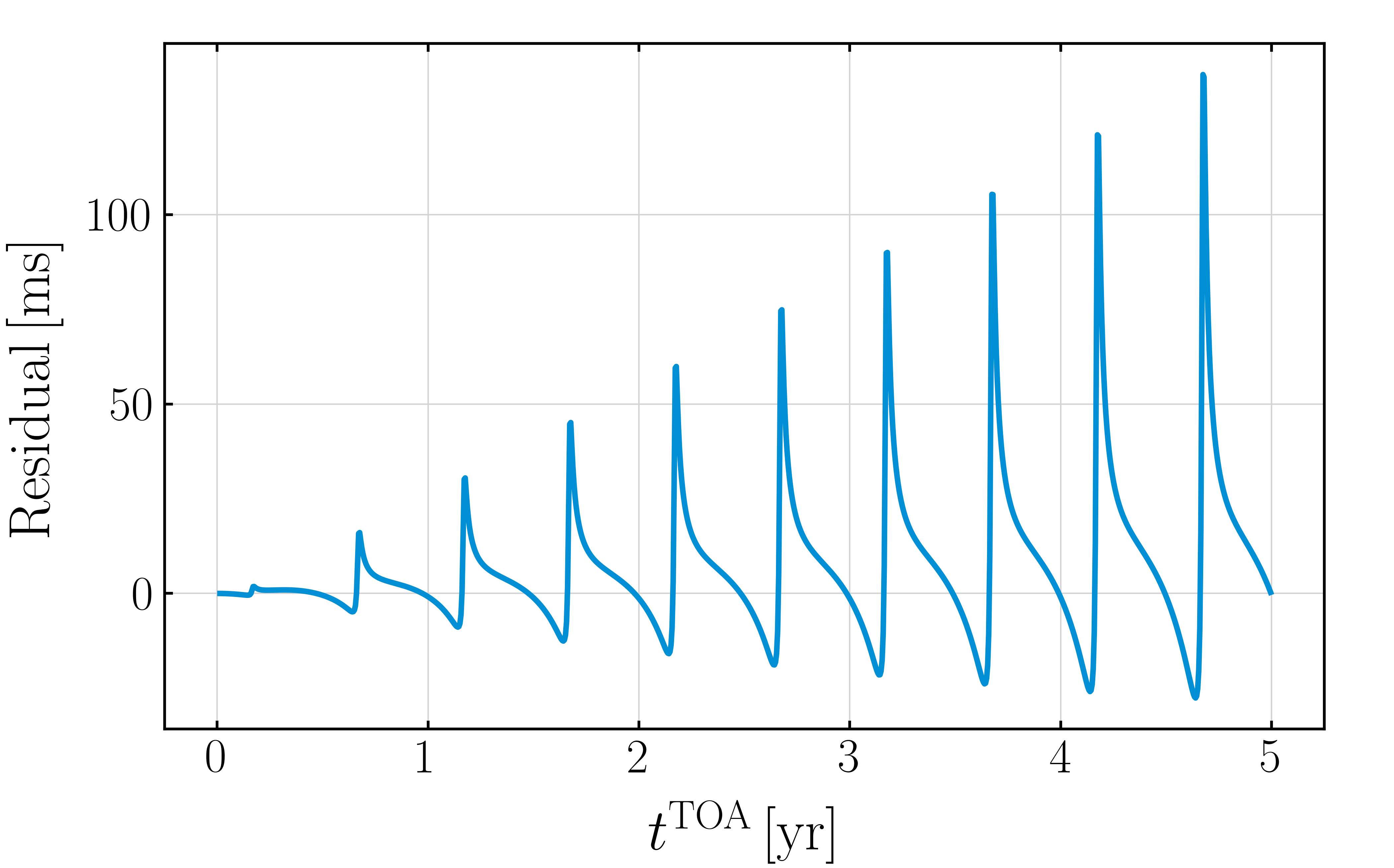}
\caption{Time residuals caused by spherical soliton with $\alpha = 0.01$ and $\beta=10^{-6}$.}
\label{fig:tr}
\end{figure*}

If we do not include the ULDM's effect in our timing model, the extra time residual in $t^{\text{TOA}}$ will occur. Here we show the corresponding time residual produced by spherical soliton with $\alpha = 0.01$ and $\beta=10^{-6}$ in Fig.~\ref{fig:tr}.  

In order to better match the actual situation, we also add the proper motion of Sgr A$^*$ in our timing model. We take $\mu_\alpha = -3.2$ mas/yr and $\mu_\delta=-5.6$ mas/yr. We can see the maximum amplitude of residual could reach $\sim 100$ ms in case of $\beta = 10^{-6}$, highly exceeding the timing precision $\sigma_\text{TOA}=1 $ms. However, $\beta\sim 10^{-6}$ is predicted in our parameter estimation. The reason is that the effects of ULDM will exhibit a certain degree of parameter degeneracy with the spin-orbit (SO) coupling term and other terms, thereby degrading the estimation accuracy of $\beta$.


\section{Impacts of BH spin and orbital inclination on the parameter estimation}\label{sec:ad}
\subsection{Impacts of BH spin}
\begin{figure*} 
\includegraphics[scale=0.3]{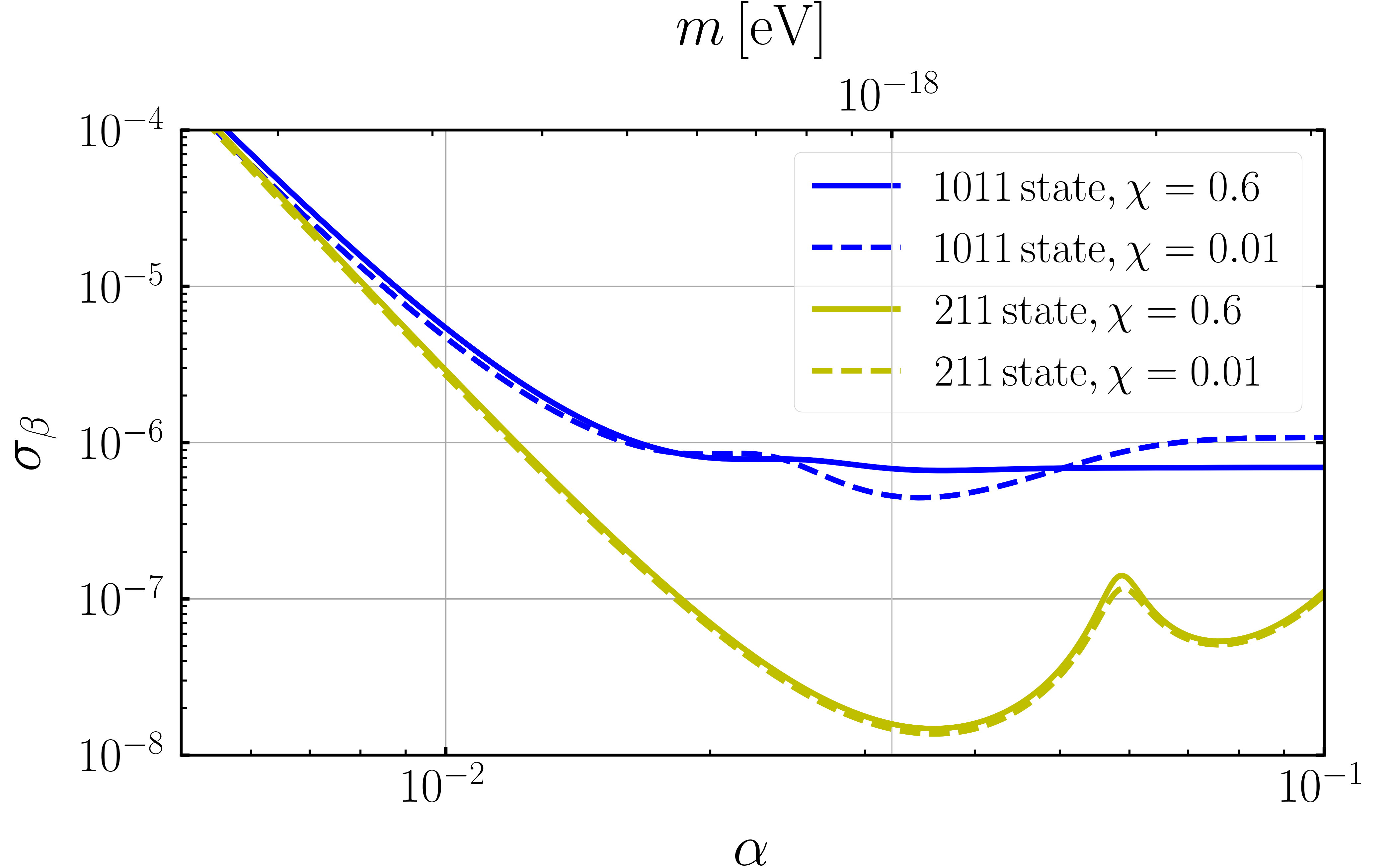}
\caption{Estimated sensitivity to $\beta$ for different spin $\chi$. For the solid (dashed) curve, the spin corresponds to 0.6 (0.01).}
\label{fig:chicom}
\end{figure*}

We select $\chi = 0.6$ in our simulation, however, the spin of BH is reduced to $\chi \sim 4\alpha/\texttt{m}$ in a saturated GA. From a theoretical standpoint, the difference in spin of $\mathcal{O}(1)$ leads to a corresponding $\mathcal{O}(1)$ change in the sensitivity to $\beta$. Our simulations further show that even for spins as small as $\chi=0.01$, the sensitivity to $\beta$ remains of the same order as in the case with $\chi=0.6$. The results are showed in Fig.~\ref{fig:chicom}. We therefore neglect the dependence of parameter estimation on the spin magnitude.
\subsection{Impacts of orbital inclination in the case of $|211\rangle$ state}
\begin{figure*} 
\includegraphics[scale=0.3]{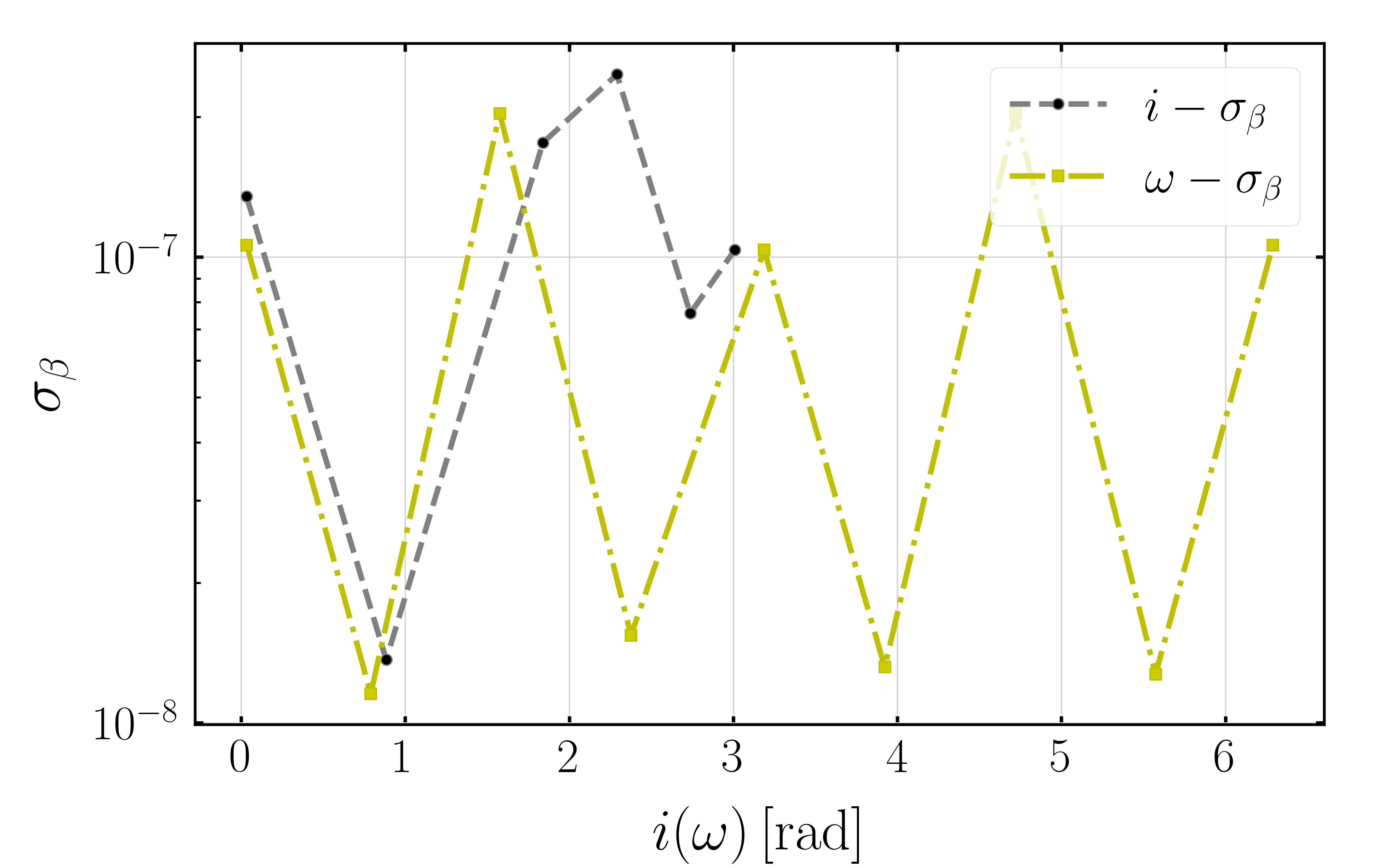}
\caption{Estimated sensitivity to $\beta$ for different inclination angle $i$ and longitude of the periastron $\omega$. For the gray (yellow) curve, the horizontal axis corresponds to $i$ ($\omega$).}
\label{fig:i}
\end{figure*}

In case of scalar $|211\rangle$ state, we draw the result that how the sensitivity of $\beta$ evolves with the inclination angle $i$ and the longitude of the periastron $\omega$ in Fig.~\ref{fig:i}. The gray line shows the variation of $\beta$ with $i$, while the yellow curve represents the dependence of $\beta$ on $\omega$.  We can conclude that the results are periodic for $\omega$ and the tighter and loosest sensitivities differ by  $\sim\mathcal{O}(10)$ for both $i$ and $\omega$.

\section{Estimation of constraints from S2 star} \label{sec:appendixC}
We estimate the constraints on $\beta$ from the observation of S2 orbit based on its periastron precession rate reported by \cite{2020S2}. The instantaneous periastron precession rate of the osculating orbit defined in Fig.~\ref{fig:frame} under a perturbing acceleration $\mathbf{F}=F_r \mathbf{e}_r+F_t \mathbf{e}_t +F_n \hat{\mathbf{L}}$ (with $\hat{\mathbf{L}}$ being the unit vector parallel to the orbital angular momentum and $\mathbf{e}_t=\hat{\mathbf{L}}\times\mathbf{e}_r$) is given by\footnote{See for example \cite{Cao:2024wby}, with the replacement $\varphi_0+\pi/2\to \omega$ (argument of periastron), $\varphi \to \theta$ (true anomaly).}
\begin{equation}
\begin{aligned}
\dot \omega = \frac{\sqrt{1-e^2}}{ae(2\pi/T)}\left\{\left[1+\frac{r}{a(1-e^2)}\right]\sin\theta\, F_t-\cos\theta\, F_r\right\}
-\cot i 
\frac{r \sin (\theta+\omega) F_n}{a^2(2\pi/T)\sqrt{1-e^2}}.
\end{aligned}
\end{equation}
The periastron shift per orbital period $T=2\pi/\sqrt{M/a^3}$ is approximately given by $\langle \dot\omega \rangle=\frac{1}{T}\int_0^T dt\,\dot\omega$, with the orbital elements in the integrand fixed to constant values. The 1PN correction \eqref{1PN} gives rise to the Schwarzschild precession:
\begin{equation}
\langle \dot\omega \rangle_\text{S}=\frac{3 M^{3/2}}{a^{5/2}(1-e^2)},
\end{equation}
(at this level of estimation, we do not consider higher PN corrections) while $\mathbf{F}=-\nabla\Phi$ gives rise to an additional contribution $\langle \dot\omega \rangle_\text{DM}$. We then derive the constraint by imposing that \cite{2020S2} $(\langle \dot \omega \rangle_\text{S}+\langle \dot\omega \rangle_\text{DM}) T \in 12.1'\times [1.1+0.19\times(-1, 1)]$, taking $M=4.3\times 10^6\,M_\odot$, $e=0.886$ and $a=1002\,\text{AU}$.  For the $|211\rangle$ state, we assume for simplicity that the BH spin is perpendicular to the orbital plane of S2.

\section{Newtonian potential of GA occupied by multiple bound eigenstates}
In our analysis, the GA is assumed to be dominated by a single eigenstate. In the more general scenario, the simultaneous occupation of multiple states (including the free states) is possible. In fact, a completely generic bound state can be expanded instantaneously as a superposition of stationary hydrogenic bound eigenstates, that fixes the instantaneous Newtonian potential. Here we briefly discuss the effects of multiple bound eigenstates, leaving a more detailed investigation to future work.

The mass density \eqref{rho_sphere} associated with the bound state \eqref{general_bound} is
\begin{equation}
\begin{aligned}
\rho_{l_*m_*}
&=\int Y_{l_*m_*}\sin\theta \,d\theta \,d\phi\,m|\psi_I|^2
\\
&=m\int Y_{l_*m_*}\sin\theta \,d\theta \,d\phi\left[\sum_{i,i'}c_{i}\,c_{i'}^*\psi^{(i)}\psi^{(i')*}\right]
\\
&=mr_\text{c}^{-3}\sum_{i,i'}c_{i}\,c_{i'}^*\,I^{i,i'}_\Omega \,e^{i\left[E^{(i')}-E^{(i)}\right] t}R_{i}(x)R_{i'}(x)
,
\end{aligned}
\end{equation}
where $i$ refers to the set of quantum numbers $\{n,l,\texttt{m}\}$, and
\begin{equation}
I^{i,i'}_\Omega \equiv \int Y_{l_*m_*}\,Y_i\,Y_{i'}^*\sin\theta \,d\theta \,d\phi.
\end{equation}

The Newtonian potential is given by Eq.~\eqref{Poisson}:
\begin{equation}\label{multi-modes_Phi}
\begin{aligned}
\Phi &=\sum_{l_*\ge 0,\, m_*=\texttt{m}'-\texttt{m}}\frac{-4\pi}{2l_*+1}
\int_0^\infty ds\,\rho_{l_*m_*}(s)\,s^2 \left[\frac{s^{l_*}}{r^{l_*+1}}\Theta(r-s)+\frac{r^{l_*}}{s^{l_*+1}}\Theta(s-r)\right]
Y_{l_* m_*}^*(\theta,\phi)
\\
&=
\sum_{l_*\ge |\texttt{m}'-\texttt{m}|}\frac{-4\pi m}{2l_*+1}
\sum_{i,i'}c_i\,c_{i'}^*\,I^{i,i'}_\Omega\,\frac{I_r^{i,i'}}{r_\text{c}}\,
Y_{l_*,\texttt{m}'-\texttt{m}}^*(\theta,\phi)\, \,e^{i\left[E^{(i')}-E^{(i)}\right] t}
,
\end{aligned}
\end{equation}
(where $\Theta(x)$ is the Heaviside unit step function) with
\begin{equation}
\begin{aligned}
I_r^{i,i'}\equiv \int_0^\infty dx\,x^2\,R_i(x)\,F_{l_*}(x)\,R_i(x),
\end{aligned}
\end{equation}
and
\begin{equation}
F_{l_*}(x) = \frac{x^{l_*}}{x_*^{l_*+1}}\Theta(x_*-x)+\frac{x_*^{l_*}}{x^{l_*+1}}\Theta(x-x_*),\quad x_*=x/r_\text{c}.
\end{equation}
Note that the structure of Eq.~\eqref{multi-modes_Phi} is almost identical with the bound-bound transition matrix element of GA in the presence of a companion\footnote{This connection is natural, since (in the nonrelativistic regime) it is the Newtonian potential of the cloud that is responsible for the backreaction of the companion-induced GA transitions, although the backreaction is typically analyzed using the flux-balance equations.} \cite{Cao:2024wby}, where the dipole term ($l_*=1$) is modified in the comoving frame of the central object due to its acceleration. In the present context, the dipole term in Eq.~\eqref{multi-modes_Phi} accelerates the central object, since from Eq.~\eqref{Poisson} the gravitational acceleration is given by
\begin{equation}
\begin{aligned}
\frac{-\nabla\Phi}{\frac{4\pi}{2l_*+1}} &=\sum_{l_*m_*}\left\{Y_{l_*m_*}\partial_r\left[\frac{q_{l_*m_*}}{r^{l_*+1}}+r^{l_*}p_{l_*m_*}\right]\mathbf{e}_r+\left[\frac{q_{l_*m_*}}{r^{l_*+2}}+r^{l_*-1}p_{l_*m_*}\right]\left(\mathbf{e}_\theta\partial_\theta+\mathbf{e}_\phi\frac{\partial_\phi}{\sin\theta}\right)Y_{l_*m_*}\right\},
\end{aligned}
\end{equation}
while at the center,
\begin{equation}
\begin{aligned}
\lim_{r\to 0}\partial_r\left[\frac{q_{l_*m_*}}{r^{l_*+1}}+r^{l_*}p_{l_*m_*}\right] & = \lim_{r\to 0}l_* r^{l_*-1}p_{l_*m_*} = p_{1m_*}\,\delta_{l_*,1}
,
\\
\lim_{r\to 0}\left[\frac{q_{l_*m_*}}{r^{l_*+2}}+r^{l_*-1}p_{l_*m_*}\right] & = \lim_{r\to 0} r^{l_*-1}p_{l_*m_*} = p_{1m_*}\,\delta_{l_*,1},
\end{aligned}
\end{equation}
we thus obtain
\begin{equation}
\begin{aligned}
-\nabla\Phi|_{\mathbf{r=0}} & =\frac{4\pi}{3}\sum_{|m_*|\le 1}p_{1m_*}\left(\mathbf{e}_r+\mathbf{e}_\theta\partial_\theta+\mathbf{e}_\phi \frac{\partial_\phi}{\sin\theta}\right)Y_{1m_*}
\\
&=\sqrt{\frac{4\pi}{3}}\sum_{|m_*|\le 1}p_{1m_*}\,\boldsymbol{\xi}^{m}
,
\end{aligned}
\end{equation}
with $\boldsymbol{\xi}^0=\mathbf{e}_z$ and $\boldsymbol{\xi}^{\pm 1}=\mp (\mathbf{e}_x\pm i \mathbf{e}_y)/\sqrt{2}$.

As an illustration, the density distribution of a two-level system is
\begin{align}
\rho &=\frac{M_\text{c}(t=0)}{r_\text{c}^3}|C_1\psi^{(1)}+C_2\psi^{(2)}|^2
\\
&=\frac{M_\text{c}(t=0)}{r_\text{c}^3}\left[ |C_1|^2|\psi^{(1)}|^2+|C_2|^2|\psi^{(2)}|^2+C_1C_2^*\psi^{(1)}\psi^{(2)*}+C_1^*C_2\psi^{(1)*}\psi^{(2)}
\right]
\\
&\equiv \frac{M_\text{c}(t=0)}{r_\text{c}^3}g(x,\theta,\phi),
\end{align}
here the normalization is taken to be $|C_1|^2+|C_2|^2=1$ at $t=0$. For convenience we choose $\psi^{(i=1,2)}$ to be time-independent, with $C_i=A_i(t) e^{i\chi_i(t)}$ and $A,\chi\in\mathbb{R}$. As a concrete example, for $|1\rangle=|211\rangle$ and $|2\rangle=|21,-1\rangle$,
\begin{equation}
g=\frac{1}{64 \pi } e^{-x} x^2 \sin ^2\theta  \left[A_1^2+A_2^2-2 A_1A_2 \cos (\chi_1-\chi_2+2 \phi )\right],
\end{equation}
correspondingly the Newtonian potential is
\begin{equation}
\begin{aligned}
\frac{\Phi}{\frac{M_\text{c}(t=0)}{r_\text{c}}} &=\frac{e^{-x}}{16 x^3} \left(x^5+6 x^4+24 x^3+72 x^2+144 x-144 e^x+144\right) \\
&\qquad \times \sin ^2\theta \left[A_1^2+A_2^2-2 A_1 A_2 \cos (\chi_1-\chi_2+2 \phi )\right]
\\
&\quad -\frac{e^{-x}}{4 x^3}\left[x^3+8 x^2+4 e^x \left(x^2-6\right)+24 x+24\right] \left(A_1^2+A_2^2\right).
\end{aligned}
\end{equation}
There are only $l_*= |m_*|=2$ terms. For the purely hydrogenic GA, $\chi_{211}=-E^{(2)}t=\chi_{21,-1}$, while $A_{211},A_{21,-1}$ are constant, the potential is thus stationary but not axially symmetric. The spin of the central object can break the degeneracy of this pair of hyperfine levels, and consequently the potential felt by a companion in circular orbit with angular velocity $\Omega$ depends on $\chi_1-\chi_2+2\phi=\left\{2\Omega-\left[E^{(211)}-E^{(21,-1)}\right]\right\}t$, which is stationary only at resonance, i.e., when $\Omega={\Delta E}/{\Delta \texttt{m}}=\left[E^{(211)}-E^{(21,-1)}\right]/2$.


\end{document}